\documentclass[twocolumn]{aastex631}
\usepackage{amssymb,amsmath,amsfonts}
\usepackage{graphicx}
\usepackage{xcolor}
\usepackage{epsfig}
\usepackage{verbatim}
\usepackage{color}
\usepackage{mdframed}
\usepackage{soul} 
\usepackage{braket}
\usepackage{enumitem}
\usepackage[percent]{overpic}
\usepackage{float}
\usepackage{array,booktabs,tabularx}
\renewcommand{\arraystretch}{1.6}
\usepackage{soul}
\include{eps2pdf}
\usepackage{fontawesome}
\usepackage{sidecap}
\usepackage{multirow}
\usepackage[flushmargin,bottom]{footmisc}
\def\apjl{Astrophys.\ J.\ Lett.}

\def\aap{Astron.\ Astrophys.}

\def\apjs{Astrophys.\ J. Supp.}

\def\d{\partial}

\def\ln{{\rm ln}}

\def\be{\begin{equation}}
\def\ee{\end{equation}}
\def\bea{\begin{eqnarray}}
\def\eea{\end{eqnarray}}
\def\ba{\begin{align}}

\def\bi{\begin{itemize}}
\def\ei{\end{itemize}}

\shorttitle{Modified Gravity and Dark Energy from Line Intensity Mapping}
\shortauthors{Moradinezhad Dizgah \& Bellini et al.}

\begin{document} 

\title{Probing Dark Energy and Modifications of Gravity \\ \vspace{.1in} with Ground-Based Millimeter-Wavelength Line Intensity Mapping \\ \vspace{.1in}} 

\author{Azadeh Moradinezhad Dizgah}
\affiliation{D\'epartement de Physique Th\'eorique,
Universit\'e de Gen\`eve, 24 quai Ernest Ansermet, 1211 Gen\`eva 4, Switzerland}
\author{Emilio Bellini$^{\ast}$}
\affiliation{INFN, National Institute for Nuclear Physics, Via Valerio 2, I-34127 Trieste, Italy}
\affiliation{IFPU, Institute for Fundamental Physics of the Universe, via Beirut 2, 34151 Trieste, Italy}
\affiliation{SISSA, International School for Advanced Studies, Via Bonomea 265, 34136 Trieste, Italy}
\author{Garrett~K.~Keating}
\affiliation{Center for Astrophysics, Harvard \& Smithsonian, 60 Garden Street, Cambridge, MA 02138, USA}

\email{$^{\ast}$ emilio.bellini@ts.infn.it}

\begin{abstract}
Line intensity mapping (LIM) can provide a powerful means to constrain the theory of gravity and the nature of dark energy at low and high redshifts by mapping the large-scale structure (LSS) over many redshift epochs. In this paper, we investigate the potential of the next generation ground-based millimeter-wavelength LIM surveys in constraining several models beyond $\Lambda$CDM, involving either a dynamic dark energy component or modifications of the theory of gravity. Limiting ourselves to two-point clustering statistics, we consider the measurements of auto-spectra of several CO rotational lines (from J=2-1 to J=6-5) and the [CII] fine structure line in the redshift range of $0.25<z<12$. We consider different models beyond $\Lambda$CDM, each one with different signatures and peculiarities. Among them, we focus on Jordan-Brans-Dicke and axion-driven early dark energy models as examples of well-studied scalar-tensor theories acting at late and early times respectively. Additionally, we consider three phenomenological models based on an effective description of gravity at cosmological scales. 
We show that LIM surveys deployable within a decade (with $\sim 10^8$ spectrometer hours) have the potential to improve upon the current bounds on all considered models significantly. The level of improvements range from a factor of a few to an order of magnitude.
\end{abstract}


\section{Introduction}

Deciphering the physics responsible for the observed current accelerated expansion of the Universe -- and whether it is driven by a cosmological constant or a dynamical dark energy (DE) component, or results from modifications to Einstein’s theory of gravity (MG) -- is one of the key open questions in modern cosmology. Shedding light on this question is the driving science case for several upcoming galaxy surveys, such as Euclid \citep{Amendola:2016saw} and DESI \citep{DESI:2019jxc}, which will map the large-scale structure (LSS) at redshifts $z \lesssim 2$ by measuring shape and clustering of galaxies. In the standard cosmological model (among others), DE is a sub-dominant component of the Universe's energy density at redshifts $z\gtrsim 1$. On the other hand, models with an increased density of DE density at higher redshifts have been also considered in the literature. Examples include models with early-DE (EDE) component prior to recombination (e.g., \cite{Karwal:2016vyq}) that has attracted significant attention in recent years in the context of the Hubble tension (e.g., \cite{Poulin:2018cxd, Ivanov:2020ril,Hill:2020osr, Simon:2022adh}). In addition, there are also models exhibiting a ``tracking behavior'' at late-times: the equation of state of DE $w(z)$ tracks the dominant component of the universe in that particular epoch ($w\simeq0$ during matter domination and $w\simeq-1$ for $z\lesssim 1$), so that their energy density at high-redshifts is non-negligible (e.g., \cite{Urena-Lopez:2000ewq,Bassett:2002qu,Rakhi:2009qf}). Beyond constraining possible modifications to the theory of gravity that can explain the current accelerated expansion, testing general relativity (GR) on cosmological scales is of significant interest and is highly complementary to constraints from Solar System, and binary systems \citep{Berti:2015itd}. Modifications to GR and a dynamic DE component can both modify the background expansion of the Universe and the growth of structure. Therefore, statistical properties of the LSS enable us to constrain the nature of gravity and DE.

LIM is emerging as a powerful probe of the LSS. In contrast to galaxy surveys that map the LSS by resolving individual galaxies, LIM relies on measuring cumulative spectral-line emission from ensemble of galaxies or intergalactic medium \citep{Pritchard:2011xb,Kovetz:2017agg}. The measurements of spatial fluctuations in the intensity of the spectral lines together with their precisely measured frequencies  enable us to measure the expansion history and growth of structure over extended redshift epochs and a wide range of scales, largely inaccessible to traditional galaxy surveys. In addition to 21cm line of hyperfine transition of neutral Hydrogen, LIM at mm/submm wavelengths, targeting spectral emission lines with rest-frame wavelengths in the far infrared (e.g., rotational lines of carbon monoxide CO(J$\rightarrow$J-1) \citep{Righi:2008br,Lidz:2011dx,Breysse:2014uia,Mashian:2015his,Li:2015gqa,Padmanabhan:2017ate} and fine structure of ionized carbon [CII]) \citep{Gong:2011ts,Silva:2014ira,Pullen:2017ogs,Padmanabhan:2018yul}, has recently attracted a growing attention both in the context of galaxy/star formation \citep{Gong_2011,Kovetz:2019uss,Bernal:2022jap} and cosmology \citep{Karkare:2018sar,Creque-Sarbinowski:2018ebl,MoradinezhadDizgah:2018zrs,MoradinezhadDizgah:2018lac,Liu:2020izx,Gong:2020lim,Bernal:2020lkd,Bernal:2021ylz,Schaan:2021hhy,Schaan:2021gzb,Scott:2022fev}. The wide array of pathfinder experiments such as CCAT-Prime \citep{CCAT-prime:2022qkj}, COMAP \citep{Cleary:2021dsp}, CONCERTO \citep{CONCERTO:2020ahk}, COPSS \citep{Keating:2016pka}, EXCLAIM \citep{Ade:2019ril}, mmIME \citep{Keating:2020wlx}, SPT-SLIM \citep{Karkare:2021ryi}, TIME \citep{TIME}, are paving the way for future wide-field surveys capable of providing high-precision constraints on cosmology by providing robust detection of the line intensity power spectrum and constraining the astrophysical model dependencies of the signal. Establishing the science cases for future surveys and determining optimized survey strategies to achieve relevant theoretical thresholds is of uttermost importance. 

Building upon our previous works on investigating the potential of mm-wavelength LIM in constraining primordial non-Gaussianity \citep{MoradinezhadDizgah:2018zrs,MoradinezhadDizgah:2018lac}, and properties of neutrinos and light relics \citep{MoradinezhadDizgah:2021upg}, in this paper, we study the prospects for constraining DE and MG.  Contrary to upcoming spectroscopic galaxy surveys mapping the LSS at redshifts $z \lesssim 2$, the broad redshift coverage of LIM observations enable us to map the background expansion and growth of structure both in the matter and DE domination epochs. As such, LIM surveys offer the possibility of testing the nature of DE or MG in the regime that has not been probed before. Furthermore, mapping the LSS on ultra large scales allows searching for horizon-scale imprints of possible modifications of GR (see e.g. \cite{Sailer:2021yzm,Berti:2021ccw,Casas:2022vik,Scott:2022fev} for recent studies on constraining beyond $\Lambda$CDM models with 21cm and mm-wavelength LIM).

For this work, we consider a ground-based survey capable of probing six rotational lines of CO and [CII] over the redshift range of $0<z<12$, assuming the same survey characterizations as those used in \cite{MoradinezhadDizgah:2021upg,Karkare:2022bai}. We analyze several beyond $\Lambda$CDM models, falling into two broad categories of {\it covariant} and {\it effective description} models. More specifically, we consider the Jordan-Brans-Dicke (JBD) theory of gravity and EDE driven by an ultra-light axion \cite{Poulin:2018cxd} as examples of \textit{covariant} models. On the other hand, within the {\it effective description}, we consider the Chevallier–Polarski–Linder (CPL) parameterization for the DE equation of state and two models for the effective description of luminal Horndeski theories. Lastly, in a hybrid approach, we consider an effective description of ShiftSymmetric Horndeski models imposing theoretical priors that ensure the consistency with the the underlying {\it covariant theory}. This selection of models enables us to explore phenomenologically rather broad regions of the parameter space of viable models to determine the potential of future LIM surveys in constraining beyond $\Lambda$CDM scenarios. 

 In our forecasts, we use \textsc{limFisher} package, which will be publicly released upon publication of this paper. This code is written in C language and is designed to perform forecasts for LIM power spectrum. It utilizes \textsc{CLASS} Boltzmann solver \citep{Blas:2011rf} {\faGithub}\footnote{\url{https://github.com/lesgourg/class_public}}, as well as its extensions for beyond $\Lambda$CDM models; In particular, we use the publicly available \textsc{CLASS\_EDE} code \citep{Hill:2020osr} {\faGithub}\footnote{\url{https://github.com/mwt5345/class_ede}} for EDE model, and \textsc{hi\_class} code \citep{Zumalacarregui:2016pph,Bellini:2017avd,Bellini:2019syt} {\faGlobe}\footnote{\url{http://hiclass-code.net}} for other models, including JBD, shift-symmetric Horndeski, and effective description of DE models. We only account for the imprints of the considered models at the linear level, as implemented in the Boltzmann codes, consistently neglecting non-linear scales.

The rest of the paper is organized as follows; in \S \ref{sec:th_MG_DE}, we review the theoretical models that we will consider, summarizing the model parameters and current bounds, as well as the imprints on the LSS. In \S \ref{sec:linePS}, we describe the model of the power spectrum of line intensity fluctuations that we use in our forecasts. In \S \ref{sec:fisher}, we summarize the Fisher forecast methodology, and in \S \ref{sec:survey}, we describe the survey specifications assumed in the forecasts. In \S \ref{sec:res}, we present the results of the forecasts, and finally in \S \ref{sec:conc}, we draw our conclusions.

\section{Review of Theoretical Models Beyond $\Lambda$CDM} \label{sec:th_MG_DE}

In this section, we first lay out the theoretical framework for selecting the models to study, and then review the main features of each of the models that we consider and their connection to the general framework, as well as the imprints of the model on the background evolution and the growth of structures.   

We perform our analysis following two complementary approaches. To study alternative theories of gravity, it is common either to specify a covariant Lagrangian, in the following referred to as \textit{covariant approach} or to build an \textit{effective description} that encapsulates the effects of the additional degree of freedom on the evolution of perturbations on a FLRW background. The two approaches offer different advantages, and we use both to explore a meaningful portion of the parameter space of viable cosmological models.

Our starting point for the \textit{covariant approach} is the following Lagrangian 
\begin{align}
  &\mathcal{L} = G_2\left(\phi,\,X\right) - G_3\left(\phi,\,X\right)\,\box\phi + G_4\left(\phi\right)\,R + \mathcal{L}_m\,,\label{eq:lag_cov}
\end{align}
where $\mathcal{L}_m$ is the matter Lagrangian, $G_{2,\,3}$ are functions of a scalar field $\phi$, and its canonical kinetic term $X\,\equiv\,-\phi^{;\mu}\phi_{;\mu}/2$, and $G_4$ is a function of the scalar field alone. This describes the sub-class of Horndeski models \citep{Horndeski:1974wa,Deffayet:2009mn,Kobayashi:2011nu} with the additional requirement that the speed of gravitational waves (GW) has to be equal to the speed light \citep{Baker:2017hug,Creminelli:2017sry,Ezquiaga:2017ekz,Sakstein:2017xjx}. This is why the $G_4$ function depends only on the scalar field and not on its derivatives; a derivative coupling, allowed in the full Horndeski framework, would modify the speed of gravitational waves. Our choice of this subclass of Horndeski models is driven by the near-simultaneous observation of the GW signal (GW17081) and the gamma-ray burst (GRB 170817A) emitted from the merger of a binary neutron stars system, which established that the GW speed has to be equal to the speed of light within one part in $10^{15}$ \citep{LIGOScientific:2017zic}. Upon fixing the functional form of $G_{2,\,3,\,4}$ in Eq.~(\ref{eq:lag_cov}) and setting the initial conditions for the scalar field, it is possible to solve the background expansion history and the perturbations consistently. However, for non-standard choices of $G_{2,\,3,\,4}$, it is difficult to predict the evolution of the background and the perturbations. 

An alternative approach that is not specific to a particular (covariant) Lagrangian is to utilize the ``effective description'' of dark energy, often denoted as Effective Field Theory for DE (EFTofDE). Here the information content of Eq.~(\ref{eq:lag_cov}) can be compressed into a few meaningful functions of time \citep{Gubitosi:2012hu,Gleyzes:2013ooa,Bellini:2014fua}. Up to linear order in perturbation theory (PT), the dynamics of Eq.~(\ref{eq:lag_cov}) can be fully described in terms of four\footnote{Note that here we are considering a sub-class of Horndeski. The full Horndeski theory requires five functions.} functions of time\footnote{The derivative of the effective Planck mass $M_*^2$ is commonly written as $\alpha_M\,\equiv\,d\ln M_*^2/d\ln a$.}
\begin{align}
    &\{w,\,\alpha_{\rm K},\,\alpha_{\rm B},\,M_*^2\}\,,\label{eq:alphas}
\end{align}
where $w\equiv p_{\rm DE}/\rho_{\rm DE}$ is the equation of state of DE that contributes to the expansion history of the universe, while the remaining three functions affect only the evolution of the perturbations. In particular, $\alpha_{\rm K}$ is the standard kinetic term of perfect fluids, $\alpha_{\rm B}$ is the kinetic term arising from the coupling of the derivatives of the scalar field with the ones of the metric, and $M_*^2$ is an effective Planck mass squared characterizing the change in the strength of gravity\footnote{For more details on the properties of these functions, see e.g.,~\cite{Bellini:2014fua}}. One of the key advantages of this approach is that each one of these functions has a clear physical meaning, and they describe particular classes of models. In this scenario, one has the freedom of parameterizing these functions of time, keeping in mind that in $\Lambda$CDM they reduce to
\begin{align}
    &\{w=-1,\,\alpha_{\rm K}=0,\,\alpha_{\rm B}=0,\,M_*^2=1\}\,.\label{eq:alphas_lcdm}
\end{align}

The two approaches to modeling alternative theories of gravity are to some extent equivalent, but each one has its own peculiarities and is preferable in specific situations. The \textit{covariant approach} is fully consistent, i.e.\ for a single theory it is possible to describe gravitational interactions in any regime. Still, it is non-trivial to jump from one model to the other, and one needs some clever intuition to choose $G_{2,\,3,\,4}$. On the other hand, the \textit{effective description} is ideal for investigating general classes of models at once. It is closer to observations but can lead to non-physical theories. Indeed, while any choice of the $G_{2,\,3,\,4}$ functions has a corresponding time evolution of $\{w,\,\alpha_{\rm K},\,\alpha_{\rm B},\,M_*^2\}$, it is not guaranteed that any time evolution of $\{w,\,\alpha_{\rm K},\,\alpha_{\rm B},\,M_*^2\}$ is the representation of a particular choice of $G_{2,\,3,\,4}$. The choice of approach depends on the needs of the analysis.

\subsection{Jordan-Brans-Dicke}

First, we study the constraints on the Jordan-Brans-Dicke (JBD) theory of gravity \citep{Brans:1961sx}, the workhorse of modified gravity models, extensively studied in the literature, e.g.~\cite{Lima:2015xia,Umilta:2015cta,Ballardini:2016cvy,Joudaki:2020shz}. The JBD theory adds a new scalar d.o.f.\ with respect to GR, that is non-minimally coupled to the metric. Following the \textit{covariant approach}, JBD is recovered by specifying
\begin{align}
    &G_2\, =\, \frac{M_{\rm Pl}^2}{\phi} \omega_{\rm BD} X - V\left(\phi\right),\\
    &G_3\, =\, 0,\\
    &G_4\, =\, \frac{M_{\rm Pl}^2}{2}\phi\,,
\end{align}
where $V\left(\phi\right)$ is a generic potential and $\omega_{\rm BD}$ is the coupling constant of the theory. The GR limit is recovered when $\omega_{\rm BD}\rightarrow\infty$. Therefore, the accelerated expansion of the Universe can be obtained for a sufficiently flat potential and a slow-rolling scalar field.

We restrict our analysis to a constant potential, i.e.~$V\left(\phi\right) = \Lambda$ (see, e.g., \cite{Ballardini:2016cvy} for constraints on power-law potential). Since $\Lambda$ can be obtained by imposing that in a flat universe, the sum of the fractional densities has to be one, the free parameters of the theory are $\{\omega_{\rm BD},\,{\rm IC}\}$, where ${\rm IC}$ stands for initial conditions of the scalar field. The initial value for the scalar field's first derivative is irrelevant since the scalar field evolves quickly to an attractor solution irrespective of the value of the first derivative. It is possible to recast the scalar field as an effective Newton's constant using
\begin{align}\label{eq:jbd_geff}
    & {\tilde G}_{\rm eff} = \frac{G_{\rm eff}^{\rm today}}{G} = \frac{4+2\omega_{\rm BD}}{\phi\left(3+2\omega_{\rm BD}\right)}\,.
\end{align}
Then, the only two parameters that will be varied are
\begin{align}
    &\{w^{-1}_{\rm BD},\, {\tilde G}_{\rm eff}\}\,.
\end{align}

The strongest bounds on JBD gravity come from astrophysical probes. In particular, the stellar triple system PSR J0337+1715, where the inner pulsar-white dwarf binary is in orbit with another dwarf, was shown to provide the strongest bound on the market, i.e.~$\omega_{\rm BD}>1.4\times 10^5$ ($95\%$ CL) \citep{Archibald:2018oxs,Voisin:2020lqi}. The current cosmological constraints are much weaker, i.e.~$\omega_{\rm BD}>2230$ and ${\tilde G}_{\rm eff} = 0.996\pm 0.029$ ($95\%$ CL), from KiDSx2dFLenS, BOSS full shape, Planck18 and Pantheon data \citep{Joudaki:2020shz}. This is because the JBD model has no screening mechanism, which implies that the fifth force introduced by the scalar field at cosmological scales is not suppressed at small scales. Therefore, in the quest for the ultimate theory of gravity, JBD becomes less interesting than other competitors. However, it is still possible to use JBD to infer when cosmological probes will have the same constraining power as astrophysical probes. Furthermore, the JBD gravity is one of the few models that has survived the recent constraints from the neutron star merger that constrained the speed of gravitational waves. It is also worth noting that the constraints on the JBD gravity can, to some extent, give insights into the long-wavelength features of the generalized scalar-tensor theories \citep{Avilez:2013dxa}. 

At the background level in JBD gravity, the scalar field begins to evolve during the radiation-dominated era as a power-law of the scale factor. The expansion is slowed down during matter domination, and at late-times, the constant gravitational potential of the scalar field sources the accelerated expansion of the universe. The scalar field evolution depends on the value of $\omega_{\rm BD}$; the larger it is, the more frozen is it. Therefore, we expect strong degeneracies between the $\tilde G_{\rm eff}$, the present-day Hubble parameter, $H_0$, the total matter overdensity, $\Omega_m$, the amplitude of fluctuations $S_8 =  \sigma_8\sqrt{\Omega_m}$, as well as a relatively mild correlation with the total sum of neutrino masses $M_\nu$ (see \cite{Joudaki:2020shz} for recent analysis).

\subsection{Early Dark Energy driven by ultra-light axions}\label{sec:ede}

The models in which a new component acting as DE dominates the energy density of the universe briefly before recombination and rapidly decays after the last scattering have recently attracted attention in the literature as a solution to alleviate the Hubble tension between the early and late universe, as proposed originally in \cite{Karwal:2016vyq}. An EDE component increases the expansion rate, thus the value of $H_0$, and acts to shrink the sound horizon at the last scattering. We consider the phenomenological model of EDE proposed in \cite{Poulin:2018cxd}, which has been used in several recent analyses of the cosmic microwave background (CMB) and LSS data (see, e.g., \cite{Ivanov:2020ril,Hill:2020osr,Sailer:2021yzm}) and got the silver medal in the $H_0$ Olympics \citep{Schoneberg:2021qvd}. This model envisages an ultra-light axion, much lighter than fuzzy dark matter axions and displaced from the minimum of its potential at early times, as an EDE component, and can be described in the \textit{covariant approach} with
\begin{align}
    &G_2\, =\, M_{\rm Pl}^2 X - m^2 f^2 \left[1-\cos\left(\frac{\phi}{f}\right)\right]^n-\Lambda,\label{eq:ede_G2}\\
    &G_3\, =\, 0,\\
    &G_4\, =\, \frac{M_{\rm Pl}^2}{2}\,.
    \label{eq:ede_bf}
\end{align}
Here, $m$ is the axion mass, $f$ is the axion decay constant, $n$ is an exponent that generalizes the well-motivated axion-like potential (recovered by setting $n=1$) and controls the rate of dilution after the field becomes dynamical, and $\Lambda$ is a cosmological constant driving the late-times accelerated expansion of the universe. Given that the dynamics are shown to be relatively insensitive to changes of $2\lesssim n \lesssim 4.5$ \citep{Smith:2019ihp,Agrawal:2019lmo}, as done in previous analyses of \cite{Hill:2020osr} and \cite{Schoneberg:2021qvd}, we fix $n=3$. Early on, due to Hubble friction, the axion stays in its displaced position and acts as an additional DE component. Once the Hubble parameter drops below the axion's mass ($H\lesssim m$), the field starts rolling down the potential and oscillates at its minimum. To have this EDE component dissipate fast and become a subdominant energy component, near the minimum, the potential should be steeper than quadratic. The physics of this model is, therefore, governed by three parameters, 
\be
\{f_{\rm EDE},\,\log_{10}(z_c), \theta_i\}\,,
\ee
where $f_{\rm EDE} \equiv {\rm max}(\rho_{\rm EDE}/3 M^2_{\rm pl} H^2)_t$ is the maximal fractional contribution of the EDE density to the total energy density of the universe, $z_c$ is the critical redshift at which this maximum is reached, and $\theta_i=\phi_i/f$ is the normalized initial value of the scalar field such that $-\pi<\theta_i<+\pi$ \footnote{The phenomenological parameters $f_{\rm EDE}$ and $z_c$ are derived parameters obtained internally in the Boltzmann code by shooting on the original particle physics parameters $m$ and $f$.}. Once $f_{\rm EDE}$ and $z_c$ are fixed, the role of $\theta_i$ is to set the dynamics of perturbations right around $z_c$ via EDE sound speed, $c_s$. The $\Lambda$CDM limit is recovered by setting the axion mass to zero in Eq.~(\ref{eq:ede_G2}). 

This class of EDE models has been recently constrained using various combinations of datasets (e.g., \cite{Poulin:2018cxd,Smith:2019ihp,Hill:2020osr,Schoneberg:2021qvd,Simon:2022adh}). We consider the results from \cite{Simon:2022adh}, and use the best-fit values as our fiducial model and their quoted $1\sigma$ errors as a baseline for comparison with our results, 
\begin{align}
    f_{\rm EDE} &= 0.116^{+0.027}_{-0.023}\\
    \log_{10}(z_c) &= 3.83^{+0.21}_{-0.15}\\
    \theta_i &= 2.89^{+0.19}_{-0.065}\,,
\end{align}
We note that the above constraints are obtained from joint analysis of Planck (temperature, polarization, and lensing) data and BOSS full-shape galaxy power spectrum, imposing prior on $h_0$ from SH0ES local distance-ladder data. Since our forecasts only include Planck priors (from temperature and polarizations data) on $\Lambda$CDM and not on the EDE model, the comparison of LIM results with current constraints should be considered only a first exploration of the potential of LIM in constraining EDE. 

To illustrate the potential of LIM data in constraining the EDE model, it is helpful to review the imprints of an EDE component on the clustering of matter and the biased tracers. The faster expansion rate before recombination in the EDE model implies a larger acoustic sound horizon at drag time, which shifts the peaks of Baryon Acoustic Oscillations (BAO) to smaller wavenumbers compared to the $\Lambda$CDM. The presence of an EDE field also affects the dynamics of the perturbations within the comoving horizon during the epochs in which EDE is relevant, slightly suppressing the growth of perturbations around the recombination. To have an EDE model consistent with the primary CMB power spectra, some of the $\Lambda$CDM parameters must shift; $\Omega_c$ is increased to compensate for the less efficient growth of perturbations, and $n_s$ is increased since the EDE is only relevant in a short time period and its effect on perturbations is scale-dependent. These parameter shifts alter the matter power spectrum, the effect (especially on smaller scales) being more prominent at higher redshifts. This redshift dependence is also encoded in the growth rate of the structure, which is measured via the redshift-space distortions (RSD) of galaxy clustering \citep{Hill:2020osr}. In addition to modifications of the matter power spectrum, the halo/galaxy abundance and clustering can also be altered in the EDE scenario. Using N-body simulations, \cite{Klypin:2020tud} showed that the halo mass function in the EDE model is enhanced compared to $\Lambda$CDM model, i.e., the EDE model predicts more halos, with the enhancement being substantially more significant at higher redshifts and higher mass halos. Furthermore, halos of a given mass were shown to form earlier and  have higher concentrations \footnote{Despite the enhanced clustering of matter fluctuations in the EDE model resulting from a higher value of $\sigma_8$, these results did not find higher clustering of halos in EDE model.}.

Apart from mapping a larger number of modes and hence having a lower cosmic variance, access to the wide range of spatial scales and redshifts makes LIM particularly powerful for constraining the EDE scenario. In terms of the scales, measuring the line power spectrum from horizon to nonlinear scales allows for constraints on the scale-dependent effect of the EDE \citep{Hill:2020osr}. Access to large scales is crucial for models of EDE with lower critical redshifts since the growth suppression is pushed to later times, and thus affects correspondingly larger scales. For example, the EDE that peaks after recombination (not a viable resolution to Hubble tension) suppresses the linear matter power spectrum in a scale-independent way on small scales and can only be probed via its effect on large scales \citep{Sailer:2021yzm}. Since the nonlinear gravitational evolution tends to reduce the difference between the EDE and $\Lambda$CDM models, probing higher redshifts is advantageous. Moreover, being sensitive to the halo mass function, the line signal picks up the imprint of the EDE on the halo abundance, which was shown to be substantial at high redshifts \citep{Klypin:2020tud}.

\subsection{Standard CPL parametrization of Dark Energy}\label{sec:cpl}

A simple and widely used description for DE models is the CPL parametrization \citep{Chevallier:2000qy,Linder:2002et}. It falls in the \textit{effective description} category, and it assumes the presence of a DE fluid with the equation of state
\begin{align}
    w = w_0 + w_a (1-a),
\end{align}
where $w_0$ is the value of the DE equation of state today, and $w_a$ modulates its time dependence, with $a$ being the scale factor. This fluid is a minimal extension of the standard cosmological model, which is recovered by setting $w_0=-1$ and $w_a=0$. The role of $w$ is only to modify the expansion rate of the universe, and it (mildly) affects the evolution of the perturbations only through a modified Hubble parameter (in the next Sections, we will present present models that have intrinsic additional freedom for the perturbations). Recent constraints from CMB, BAO, and Supernovae \citep{Planck:2018vyg} show that these parameters are compatible with the standard $\Lambda$CDM model, i.e. 
\begin{align}
    &w_0 = -0.957\pm 0.080,\\
    &w_a = -0.29^{+0.32}_{-0.26}.
\end{align}
In our forecasts we set the fiducial values of $w_0=-1$ and $w_a=0$ corresponding to $\Lambda$CDM. LIM can provide tight constraints on the CPL model by measuring the expansion history over a wide redshift range, thus, is expected to tightly constrain redshift-dependence of DE equation of state, i.e., the $w_a$ parameter.  

\subsection{Shift-symmetric Horndeski Models}\label{sec:shift_symm}

To build a very general, yet tractable, MG theory from Eq.~(\ref{eq:lag_cov}), it is possible to assume that the scalar field is slowly evolving on a cosmological background and Taylor expand the $G_{2,\,3,\,4}$ functions for small $X$. To further compress the parameter space of the theory, we place constraints only on shift-symmetric models, i.e.\ models invariant under $\phi\rightarrow\phi+c$ (where $c$ is some constant). Shift-symmetric theories are interesting because radiative corrections are parametrically suppressed around (quasi) de Sitter backgrounds \citep{Luty:2003vm,Nicolis:2008in}. Following the \textit{covariant approach}, a shift-symmetric version of Eq.~(\ref{eq:lag_cov}) expanded up to quadratic order in $X$ looks like
\begin{align}
    &G_2\, =\, c_{01}X + \frac{c_{02}}{\Lambda_2^4} X^2\label{eq:G2_ss},\\
    &G_3\, =\, -\frac{1}{\Lambda_3^3} \left(d_{01}X + \frac{d_{02}}{\Lambda_2^4} X^2\right),\label{eq:G3_ss}\\
    &G_4\, =\, \frac{M_{\rm Pl}^2}{2}\,,\label{eq:G4_ss}
\end{align}
where $\{c_{01}, c_{02}, d_{01}, d_{02}\}$ are parameters of the theory, and commonly it is assumed that $\Lambda_2^4=M^2_{\rm Pl} H_0^2$ and $\Lambda_3^3=M_{\rm Pl} H_0^2$ to ensure that all interactions have $\mathcal O(1)$ contributions to present-day cosmological background evolution. This theory is a rather minimal, yet rich four-parameter extension of $\Lambda$CDM. However, instead of constraining the parameters of this theory, we follow an alternative and more powerful approach proposed in  \cite{Garcia-Garcia:2019cvr,Traykova:2021hbr}, in which the parameter space of Eqs.~(\ref{eq:G2_ss}-\ref{eq:G4_ss}) is projected into the \textit{effective description}\footnote{\cite{Garcia-Garcia:2019cvr} is an in-depth treatment of quintessence models, while \cite{Traykova:2021hbr} shows results for the model considered here.}. In other words, they investigate how the \textit{effective description} functions evolve with time if they are obtained consistently from Eqs.~(\ref{eq:G2_ss}-\ref{eq:G4_ss}), instead of parameterizing them directly (as in the spirit of the \textit{effective description} formalism). This provides theoretical priors on the \textit{effective description} functions that ensure we are dealing with physical theories. The theoretical priors have to be interpreted as the region in the parameter space where a ``covariant theory'' will likely fall. A possible tension between data driven constraints and theoretical priors would not indicate the data have preference on non-physical theories, but rather that only few of the physical theories are in agreement with data.

As anticipated, following the results of \cite{Traykova:2021hbr}, we can safely ignore Eqs.~(\ref{eq:G2_ss}-\ref{eq:G4_ss}), and study their proposed parameters, i.e.,
\begin{align}
    &\{w_0, \,w_a,\,\hat{\alpha}_B,\,m\},
\end{align}
which come from simple parametrizations of the \textit{effective description} functions
\begin{align}\label{eq:EFTofDE}
    w &= w_0 + w_a (1-a)\\
    \alpha_{\rm B} &= \hat{\alpha}_{\rm B} {\left(\frac{a H_0}{H}\right)}^{4/m}\,.
\end{align}
It is clear that $\hat{\alpha}_{\rm B}$ represents the magnitude of $\alpha_{\rm B}$ today, the parameter $m$ modifies its dependence on the Hubble parameter $H$. The $\Lambda$CDM limit is obtained fixing $w_0=-1$, $w_a=0$ and $\hat{\alpha}_{\rm B}=0$, in which case the value of $m$ becomes irrelevant.

Combining observational constraints from CMB, BAO, RSD and Supernovae Type IA data with theoretical priors, and assuming that the scalar field is entirely responsible for the late time acceleration (case $\Lambda=0$ in \cite{Traykova:2021hbr}) \footnote{This was dubbed the ``self-accelerating'' version of the shift-symmetric model, which in general is more restrictive and easier to rule out by observational data. In this case, $G_2$ is bounded to be negative, thus, it is the form of $G_3$ that restore a positive kinetic energy and avoid ghosts.}, in \cite{Traykova:2021hbr} they obtained the following constraints on the four model parameters
\begin{align}
    w_0 &= -0.97\pm 0.03,\\
    w_a &= -0.11\pm 0.06,\\
    \hat{\alpha}_{\rm B} &= 0.6\pm 0.3,\\
    m &= 2.4\pm 0.4\, ,
\end{align}
which we will use to set the fiducial values for our forecasts and as a comparison point for results.

As already mentioned, one of the advantages of the \textit{effective description} is that different physical effects can be encapsulated by independent functions of time. These functions can thus be considered as orthogonal and treated separately. As described in \S \ref{sec:cpl}, the DE equation of state, $w$, affects primarily the expansion rate of the universe, which in turn modifies the growth of structures. Here, at the level of linear perturbations we have one more ingredient, which does not modify the expansion history, but introduces new scale dependence in the matter power spectrum. In particular, its amplitude is responsible for lowering the amplitude of the power spectrum at large scales ($k<10^{-3}h/{\rm Mpc}$) and enhancing it at small scales \footnote{See Fig. 2 of \cite{Zumalacarregui:2016pph} for an illustration of this point. Even though a different parametrization is used in this plot, its behaviour is the same as the one we are considering.}. This effect is redshift dependent, being less evident at higher redshifts, and in general depends on the time evolution of $\alpha_B$ (regulated by the parameter $m$). \vspace{0.1in}

\subsection{Effective Description of \\ Luminal Horndeski Models}

The last class of models we will investigate belongs to the \textit{effective description} class of luminal Horndeski theories. Starting from Eq.~\eqref{eq:alphas}, we parameterize the four EFT functions as
\begin{align}
    w &= w_0 + \left(1-a\right)\, w_a\\
    \alpha_{\rm K} &= a\\
    \alpha_{\rm B} &= a\\
    M_*^2 &= 1 + c_M^{(0)} - c_M^{(1)}\, \left(1-a\right) + c_M^{(2)}\, {\left(1-a\right)}^2\,.
\end{align}
The kineticity $\alpha_{\rm K}$ is largely unconstrained by cosmological data \citep{Bellini:2015xja}, due to the building blocks of $\alpha_{\rm K}$ being those of simple DE models, with at most first derivatives in the scalar field Lagrangian. As a consequence, in the sub-horizon limit, where terms with higher number of spatial derivatives become dominant, the DE fluid becomes almost negligible. We additionally fix $\alpha_{\rm B}$ in our forecast. Taking into account that its effect has been explored in \S \ref{sec:shift_symm}, and to avoid excessively enlarging the parameter space, we focus on the remaining d.o.f.~of luminal Horndeski theories, i.e., $M_*^2$. It is also important to remember that by construction the \textit{effective description} allows to separate the different physical effects, which makes it possible to study the effects of one single function at time and then draw conclusions on the general model \footnote{Strictly speaking, this is valid only when the deviation of these functions w.r.t.~to their standard $\Lambda$CDM value is $\lesssim 1$, to be able to neglect quadratic terms.}.

We keep $\alpha_{\rm K}$ and $\alpha_{\rm B}$ fixed to the evolution of the scale factor while varying the parameters related to the expansion history and Planck mass, as is similarly done in \cite{Raveri:2017qvt}, where
\begin{align}\label{eqn:raveri_params}
    &\{w_0,\,w_a,\,c_M^{(0)},\,c_M^{(1)},\,c_M^{(2)}\}\,.
\end{align}
This parametrization is a power law expansion around the present epoch up to first-order in the expansion history and second-order in the effective Planck mass. It allows a smooth GR limit (i.e. where Eq.~\eqref{eqn:raveri_params} can be expressed as $\{-1,\,0,\,0,\,0,\,0\}$), and can have rich dynamics. While the vast majority of the models proposed in the literature as late-times DE/MG kick in at $z\sim 1$, we explore the potential of LIM experiments spanning a wide range of redshifts. Thus, we choose two ad-hoc models (defined in Table~\ref{tab:models}), designed to have signatures at high-redshift, i.e., $2\lesssim z \lesssim 10$. In particular, the effective Planck mass of \textit{model1} increases smoothly with time, and its deviations w.r.t.~$1$ are $\sim10\%$ at $z=10$. \textit{model2} is even more extreme, with deviations $\sim25\%$ at $z=10$, but also a feature at $z\simeq1$ after which $M_*^2$ starts decreasing. 

Two type of information can be derived from forecasts on this model: (i) maximize the constraining power of LIM experiments on DE/MG parameters, and (ii) understand to what extent a time varying effective Planck mass can be constrained in this regime. The effective Planck mass $M_*^2$, together with its derivative $\alpha_{\rm M}$, enhances the growth of structures at large scales ($k\simeq 10^{-3} {\rm Mpc}^{-1}h$), while keeping the amplitude of the matter power spectrum almost constant at smaller scales. The effect is redshift dependent, and only models with high-redshift signatures will have the matter power spectrum modified at high-redshift.

\section{The Power spectrum of \\ \vspace{0.03in} line intensity fluctuations \\}\label{sec:linePS}

We use a simple model for the line intensity signal,in which line intensity is calculated via scaling relationships between host halo mass and the galaxies CO or [CII] luminosities. The details of the model and relevant references to previous works on modeling line power spectra can be found e.g., in \cite{MoradinezhadDizgah:2021upg}. Here we just give the final form of the line power spectrum used in our forecasts. 

The total observed power spectrum of fluctuations in a given line has three contributions: clustering, shot and instrumental noise,
\be
P_{\rm tot}(k,\mu,z) = P_{\rm clust}(k,\mu,z) + P_{\rm shot}(z) + P_N(k,z).
\ee
Here $\mu$ is the cosine of the angle between wavevector ${\bf k}$ and the line-of-sight direction. The clustering contribution (typically in units of $\mu$K$^2$ Mpc$^{-3}h^3$) is anisotropic due to the RSD and the Alcock-Paczynski (AP) effect. Assuming linear bias, on large scales the line power spectrum is given by
\begin{align}\label{eq:line_clust}
P_{\rm clust}^{\rm s}(k,\mu,z) &= \frac{H_{\rm true}(z)}{H_{\rm ref}(z)}  \left[\frac{D_{A,{\rm ref}}(z)}{D_{A,{\rm true}}(z)}\right]^2 \notag \\
&\hspace{-.2in}\times \left[1+\mu_{\rm true}^2\beta(k_{\rm true},z)\right]^2 {\rm exp}\left(-\frac{k_{\rm true}^2 \mu_{\rm true}^2 \sigma_v^2}{H^2(z)}\right),\notag \\
&\hspace{-.2in}\times \left[\bar T_{\rm line}(z)\right]^2 b_{\rm line}^2(z) P_0(k_{\rm true},z),
\end{align}
where $\bar T$ is the mean brightness temperature of the line, $b_{\rm line}$ is the linear bias of the line, $P_0$ is the linear matter power spectrum, $H$ is the Hubble parameter, $D_A$ is the comoving angular diameter distance, $\beta= f/b_{\rm line}$, and $f = d\ln D/d\ln a$ is the logarithmic growth rate of structure with $D$ being the growth factor and $a$ the scale factor. The wavenumbers and angles in the true and reference cosmologies (characterizing the AP effect) are related as 
\begin{align}\label{eq:trueref}
k_{\rm true} &= k \left[(1-\mu^2) \frac{D_{A,{\rm ref}}^2(z)}{D_{A,{\rm true}}^2(z)} + \mu^2 \frac{H_{\rm true}^2(z)}{H_{\rm ref}^2(z)}\right]^{1/2},  \nonumber \\
\mu_{\rm true} &= \frac{k \mu }{k_{\rm true}} \ \frac{H_{\rm true}(z)}{H_{\rm ref}(z)}.
\end{align}
The exponent $\sigma_v$ in Eq.~\eqref{eq:line_clust} includes both finger-of-god (FoG) effect and the intrinsic line width of individual emitters, both of which result in smearing of power on small scales.

The mean brightness temperature of the line (typically in units of $\mu$K) at redshift $z$ is given by
\begin{equation}
\bar T_{\rm line} (z)  = \frac{c^2 p_{1,\sigma}}{2k_B \nu_{\rm obs}^2} \int_{M_{\rm min}}^{M_{\rm max}} dM \frac{dn}{dM} \frac{L_{\rm line}(M,z)}{4 \pi \mathcal{D}_{L}^{2}} \left ( \frac{dl}{d\theta} \right )^{2} \frac{dl}{d\nu},
\end{equation}
where $M_{\rm min}$ and $M_{\rm max}$ are the minimum and maximum masses of the halos that host galaxies emitting in a given line, $c$ is the speed of light, $k_B$ is the Boltzmann factor, $\nu_{\rm obs}$ is the observed frequency of the line, $dn/dM$ is the halo mass function, $L_{\rm line}$ is the specific luminosity of the line-emitting galaxy located in a halo of mass $M$ at redshift $z$, and ${\mathcal D}_L$ is the luminosity distance. The terms $dl/d\theta$ and $dl/d\nu$ reflect the conversion from units of comoving lengths, $l$, to those of the observed specific intensity: frequency, $\nu$, and angular size, $\theta$. The parameter $p_{1,\sigma}$ accounts for scatter in the relations between SFR and specific luminosity with halo mass \citep{Li:2015gqa,Keating:2016pka}.

In the Poisson limit, the shot-noise contribution arising from the discrete nature of line emitters is given by 
\begin{align}\label{eq:ps_shot}
P_{\rm shot}^{\rm s}(z) = \frac{c^4 p_{2,\sigma}}{4 k_B^2 \nu_{\rm obs}^4} 
& \int_{M_{\rm min}}^{M_{\rm max}} dM  \ \frac{dn}{dM} \notag \\ 
&\hspace{-0.3in} \times{\left[\frac{L_{\rm line}(M,z)}{4 \pi \mathcal D_L^2} 
 \left ( \frac{dl}{d\theta} \right )^{2} \frac{dl}{d\nu} \right ]}^2. 
\end{align} 

We note that to obtain more realistic forecasts, the line power spectrum model used in this work should be improved to include nonlinearities in matter distribution, RSD and line biasing. Furthermore, one should account for deviations from Poisson shot noise \citep{MoradinezhadDizgah:2021dei} and marginalize over these corrections in addition to the full set of bias parameters to obtain cosmological parameters. Including more free parameters allow us to consider smaller scales in the forecasts, but at a fixed small-scale cutoff, would result weakening of the constraints compared to the linear model (see e.g., \cite{Sailer:2021yzm}). We leave a forecasts using a more complete model to future work.

\section{Analysis Methodology}\label{sec:fisher}

We use Fisher forecast formalism to determine the constraining power of a ground-based mm-wavelength intensity mapping survey, probing CO rotational lines and [CII] line. Our analysis framework is the same as what we used in our previous paper, \cite{MoradinezhadDizgah:2021upg}. Therefore, we only summarize the main features here and refer the reader to that work for further details.

Binning the survey in redshift bins of 0.1 dex, and neglecting correlations between redshift bins, for each emission line $\rm x$, the total Fisher matrix is the sum of Fisher matrices of individual redshift bins,
\be
F_{\alpha \beta}^{\rm x} = \sum_i F_{\alpha\beta}^{{\rm x},i}
\ee
with the Fisher matrix in the $i^{\text{th}}$ redshift bin given by
\begin{align}\label{eq:Fisher_single}
F_{\alpha\beta}^{{\rm x},i} &=  V_i \int_{-1}^1\int_{k_{\rm min}}^ {k_{\rm max}}  \frac{ k^2 {\rm d}k \ {\rm d}\mu  }{8 \pi^2} {\rm Var}^{-1}[P_{\rm obs}^{\rm x}(k,\mu,z)] \nonumber \\
&\times \frac{\partial P_{\rm clust}^{\rm x}(k,\mu,z_i)}{\partial { \lambda}_\alpha } \frac{\partial P_{\rm clust}^{\rm x}(k,\mu,z_i)}{\partial {\lambda}_\beta } ,
\end{align}
where $\lambda$ are the model parameters that are varied, $V_i$ is the volume of $i^{\text{th}}$ redshift bin extended between $z_{\rm min}$ and $z_{\rm max}$ and is proportional to sky coverage of the survey, $f_{\rm sky}$, while ${\rm Var}^{-1}[P_{\rm line}^{\rm x}]$ is the variance of the line power spectrum. We include the noise due to interloper lines from lower and higher redshifts in the variance of the line of interest, these interloper contributions -- typically dominated by low-redshift CO -- are assumed to be removable in the power spectrum domain  \citep[e.g.][]{Cheng2016ApJ...832..165C}.

We set the fiducial $\Lambda$CDM  parameter values to those from Planck 2018 data\footnote{Specifically, we use \textsf{base\_plikHM\_TTTEEE\_lowl\_lowE}.} \citep{Planck:2018vyg}: ${\rm ln} (10^{10}A_s) = 3.0447, n_s = 0.96589, h = 0.6732,\Omega_{\rm cdm} =0.2654,  \Omega_b = 0.04945$. The fiducial values of free parameters of each of the theoretical models considered are described in section \ref{sec:th_MG_DE}. We assume three degenerate massive neutrino species for all forecasts and fix their total mass to $M_\nu=0.06 \ {\rm MeV}$. For the finger-of-god contribution, we set the fiducial value of $\sigma_{{\rm FOG},0} = \ 250 \ {\rm km}/{\rm s}$.

We fix the line bias and mean brightness temperature to their theoretical values, accounting for their cosmology dependence. For the combined results of LIM and Planck, we use the available Planck 2018 Monte Carlo Markov chain for $\Lambda$CDM \footnote{\url{http://pla.esac.esa.int/pla/\#cosmology}} to compute the parameter covariances and the Fisher matrix, marginalizing over optical depth and assume the CMB and LIM Fisher matrices are independent so can be summed to obtain the joint constraints. A more consistent forecast including Planck and LIM data requires computing the Fisher matrices of the two observables for the same beyond $\Lambda$CDM model. The choices of minimum and maximum scales, $[k_{\rm min},k_{\rm max}]$, and the modeling of interlopers are the same as \cite{MoradinezhadDizgah:2021upg}. The list of the models that we consider in our analysis is given in Table \ref{tab:models}.

In addition to marginalizing over nonlinear biases and corrections to the Poisson shot noise, to obtain a more realistic forecast, the line mean brightness temperature, $\bar T$, should also be marginalized over. While in general considering a more extended parameter space degrades the constraints at a fixed small-scale cutoff, it also can offer advantages; At linear level, the linear bias and $\bar T$ are highly degenerate, with RSD only partially breaking the perfect degeneracy between these two parameters in isotropic power spectrum \citep{Barkana:2004zy,McQuinn:2005hk,Sailer:2022vqx}. Considering nonlinearities can further alleviate this degeneracy. As mentioned earlier, we leave consideration of more extended model of LIM power spectrum to a future work. 

\begin{table}[t]
\renewcommand{\arraystretch}{1.2}
\centering
\hspace{.3in}
\begin{tabular}{c  c }
    \hline
    Model &  Parameters \vspace{.04in}\\
    \hline
    JBD & $\omega_{\rm BD}^{-1}, {\tilde G}_{\rm eff}$ \\ 
    & $\{800, 1\}$  \\ 
    \\
    Early-DE (ULA) & $f_{\rm EDE}, \log_{10}(z_c), \theta_i$  \\ 
    & $\{0.105, 3.59, 2.71\}$\\
    \\
    CPL &  $\{w_0, w_a\}$ \\ 
    & $\{ -1, 0\}$\\
    \\
    ShiftSym-Horndeski & $w_0, w_a, \hat \alpha_B, m$ \\ 
    & $\{-0.97, -0.11, 0.6, 2.4\}$ \\
    \\
    Effective-DE & $w_0, w_a, c_M^{(0)}, c_M^{(1)}, c_M^{(2)}$ \\ 
    & model1: $\{ -1, 1, 1, 1, 0\}$ \\ 
    & model2: $\{ -1, 1, 1, -1, -2\}$ \\
    \hline
\end{tabular}\vspace{.1in}
\caption{Summary of the models considered in the Fisher forecasts, and their corresponding free parameters.}
\label{tab:models} 
\end{table}

\section{Survey Specifications}\label{sec:survey}
In this section, we briefly summarize the survey specifications that we assume in our forecasts and refer to \cite{MoradinezhadDizgah:2021upg} for a detailed description of the hypothetical next-generation mm-wave LIM survey and how those specifications translate into sensitivity estimates used in our forecasts.

The instrument noise contributing to the measured power spectrum, $P_{\rm N}$, is given in terms of per-voxel noise, $\sigma_{\rm N}$, of the original image cube as
\begin{equation}\label{eqn:noisepermode}
    P_{\rm N} = \sigma_{N}^{2}V_{\rm vox}.
\end{equation}
$V_{\rm vox}$ is the volume of individual voxels within the image cube, which can be further expressed as
\begin{equation}\label{eqn:voxelvolume}
    V_{\rm vox} = \Omega_{\rm vox}\delta\nu \left( \frac{dl}{d\theta} \right )^{2} \frac{dl}{d\nu}.
\end{equation}
Here $\Omega_{\rm vox}$ and $\delta\nu$ are the solid angle and bandwidth covered by a single voxel, respectively.

The per-voxel noise is parameterized by \textit{spectrometer hours}, $\tau_{\rm sh}$, as a proxy for the ``level-of-effort'' of an experiment. If the survey area is $\Omega_{\rm s}$, such that the number of independent pointings is given by $\Omega_{\rm s}/\Omega_{\rm vox}$, then we can express Eq.~\eqref{eqn:noisepermode} as
\begin{equation}
    P_{\rm N} = \frac{\Omega_{\rm s} \sigma_{\rm NET}^{2} \delta\nu }{{\eta_{opt}^{2}\tau_{\rm sh}}} \left( \frac{dl}{d\theta} \right )^{2} \frac{dl}{d\nu},
\end{equation}
where $\sigma_{\rm NET}$ is the noise-equivalent temperature (NET) of the detector (units of K$\cdot\sqrt{\rm s}$), and $\eta_{\rm opt}$ is the optical efficiency of the instrument. In computing $\sigma_{\rm NET}$, we assume that the spectrometer performance is close to background-limited, with the primary contributors to the instrument noise being atmospheric emission, with secondary contributions from the telescope emissivity, and various astrophysical backgrounds (including Galactic dust and the CMB). Further details on the simulated spectrometer performance can be found in \cite{MoradinezhadDizgah:2021upg}.

We define an effective instrumental noise,
\be
\tilde P_{\rm N}(k,\mu,z) =  \alpha^{-1} _{\rm max}(k,\mu)\alpha^{-1} _{\rm min}(k,\mu)P_{\rm N},
\ee 
to account for attenuation of the signal due to the finite resolution of the instrument ($\alpha_{\rm max}$) and the finite redshift and angular coverage of the survey ($\alpha_{\rm min}$). The two attenuation factors are defined in terms of the largest and smallest recoverable modes in parallel and perpendicular to line-of-sight directions. The former is determined by the redshift and angular coverage of the survey, while the latter is set by the spectral resolution of the instrument and the diameter of the aperture (see \cite{MoradinezhadDizgah:2018lac,MoradinezhadDizgah:2021upg} for the full description and explicit expressions). 

For our analysis, we consider ground-based observations from an accessible observing site with excellent proven mm-wave observing conditions, such as the South Pole or the Atacama desert. Each spectrometer is assumed to be sensitive to the entire frequency range, i.e., each spectrometer can measure all individual spectral channels simultaneously.

We vary spectrometer hours over a wide range, starting with first-detection experiments and extending to larger-scale surveys that could be fielded in the next ten years. We set the lower bound comparable to the current-generation instruments, which feature $\sim 50$ spectrometers \citep{doi:10.1117/12.2057207} and are capable of completing surveys of order $10^{5}$ spectrometer hours. As described in \cite{MoradinezhadDizgah:2021upg}, for each model, we calculate the parameter constraints while varying $f_{\rm sky}$ and spectrometer hours to determine the optimal survey strategy. We compute the constraints both with and without the addition of the Planck priors on $\Lambda$CDM parameters (only the former ones are shown in \S \ref{sec:res}).

For our analysis, we tabulate and report the results with and without interlopers -- the latter representing the optimistic view that the interloper emission can be (near-)perfectly removed by leveraging the information contained with multiple overlapping lines \citep[e.g.][]{Cheng2020ApJ...901..142C}, and the former representing the pessimistic view that no suppression of interloper ``noise'' would be possible.

\section{Results}\label{sec:res}

In this section, we present the forecasted constraints for the models discussed in \S \ref{sec:th_MG_DE}. For each model, we first describe the results of survey optimization by showing the marginalized constraints as a function of spectrometer hours and show the corresponding sky coverage required to achieve those constraints. In these plots, we show also a gray horizontal band, representing current constraints on each parameter. It is important to stress here that, when comparing the current constraints on the same parameter but for different models (e.g., $w_0$ from \S \ref{sec:cpl} and \S \ref{sec:shift_symm}) these can be different for two reason: (i) changes in the number of parameters changes, but most importantly (ii) the quoted current constraints come from different analysis which use different datasets. To illustrate the parameter degeneracies and the redshift-dependence of the constraints for each model, we show the constraints on each parameter as a function of the maximum redshift, and the marginalized constraints on parameter pairs, assuming fixed spectrometer hours and sky coverage. 
Before discussing the results of individual models, let us summarize the main trends that are evident for all of them. Considering the dependence on survey parameters, the parameter constraint improves with increasing spectrometer hours and sky coverage. Accounting for interlopers degrades the constraints. For some parameters (which we discuss below), the degradation is most significant at higher values of spectrometer hours, causing saturation of the constraining power. This implies that in those cases, the noise from the interloper dominates the thermal noise determined by the number of spectrometer hours. Furthermore, the interloper noise pushes the required sky coverage to higher values. For all models, the redshifts corresponding to the peak of star formation rate at which the amplitude of the line power spectrum is the largest ($z\sim2$) provide considerable constraining power. However, the information content of the signal at different redshifts differs in the models considered, which we will discuss in what follows. 

In table \ref{tab:optsurv_1sig_cov} and \ref{tab:optsurv_1sig_effDE} of Appendix \ref{sec:app}, we show the 1-$\sigma$ marginalized constraints for all the considered models for five stages of LIM surveys defined in terms of an approximate number of spectrometer hours. For each model and each stage, we quote the constraints with (top row) and without (bottom row) interlopes.

\subsection{Jordan-Brans-Dicke}

\begin{figure*}[t!]
    \centering
     \includegraphics[width=0.32\textwidth]{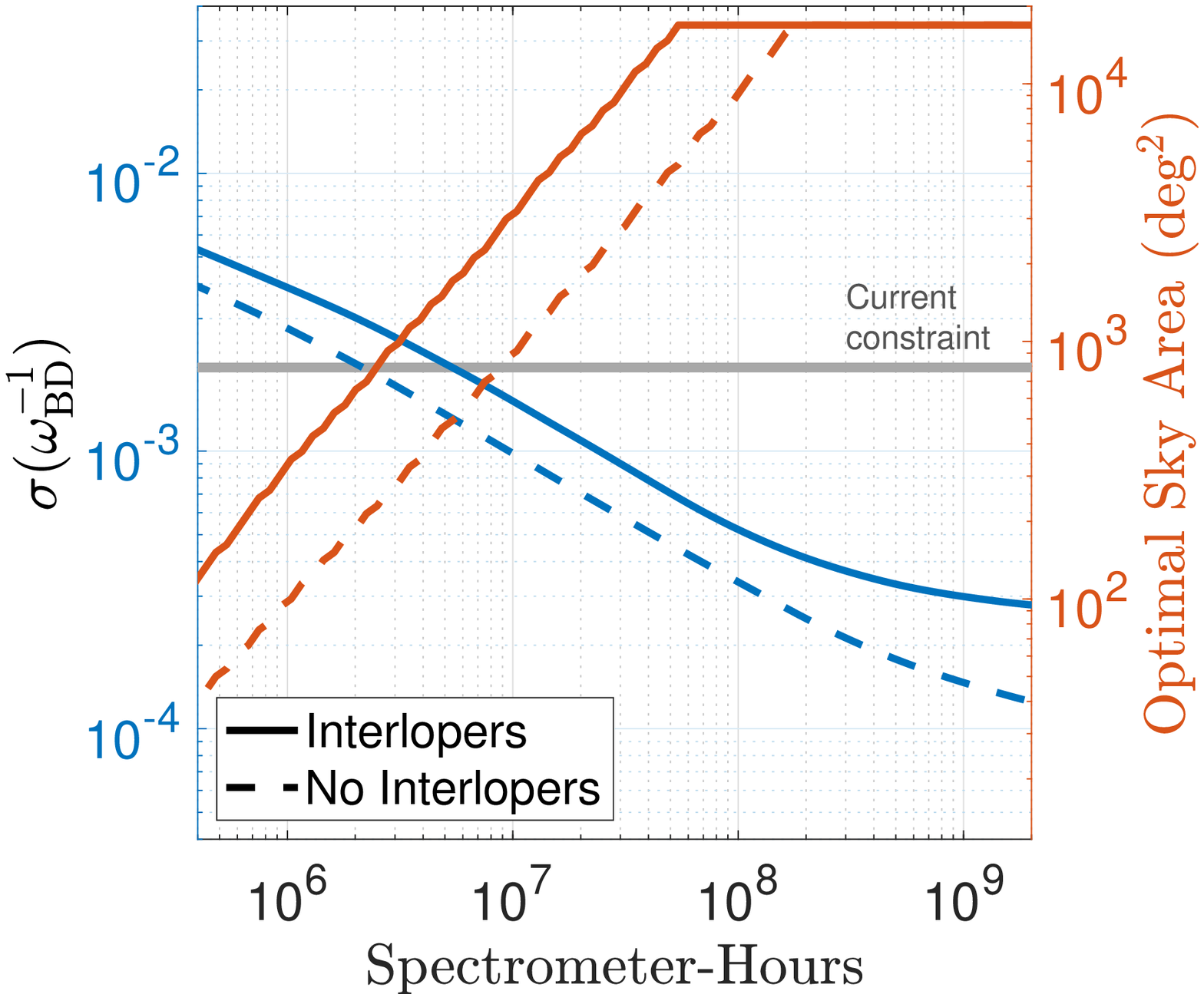} 
        \includegraphics[width=0.32 \textwidth]{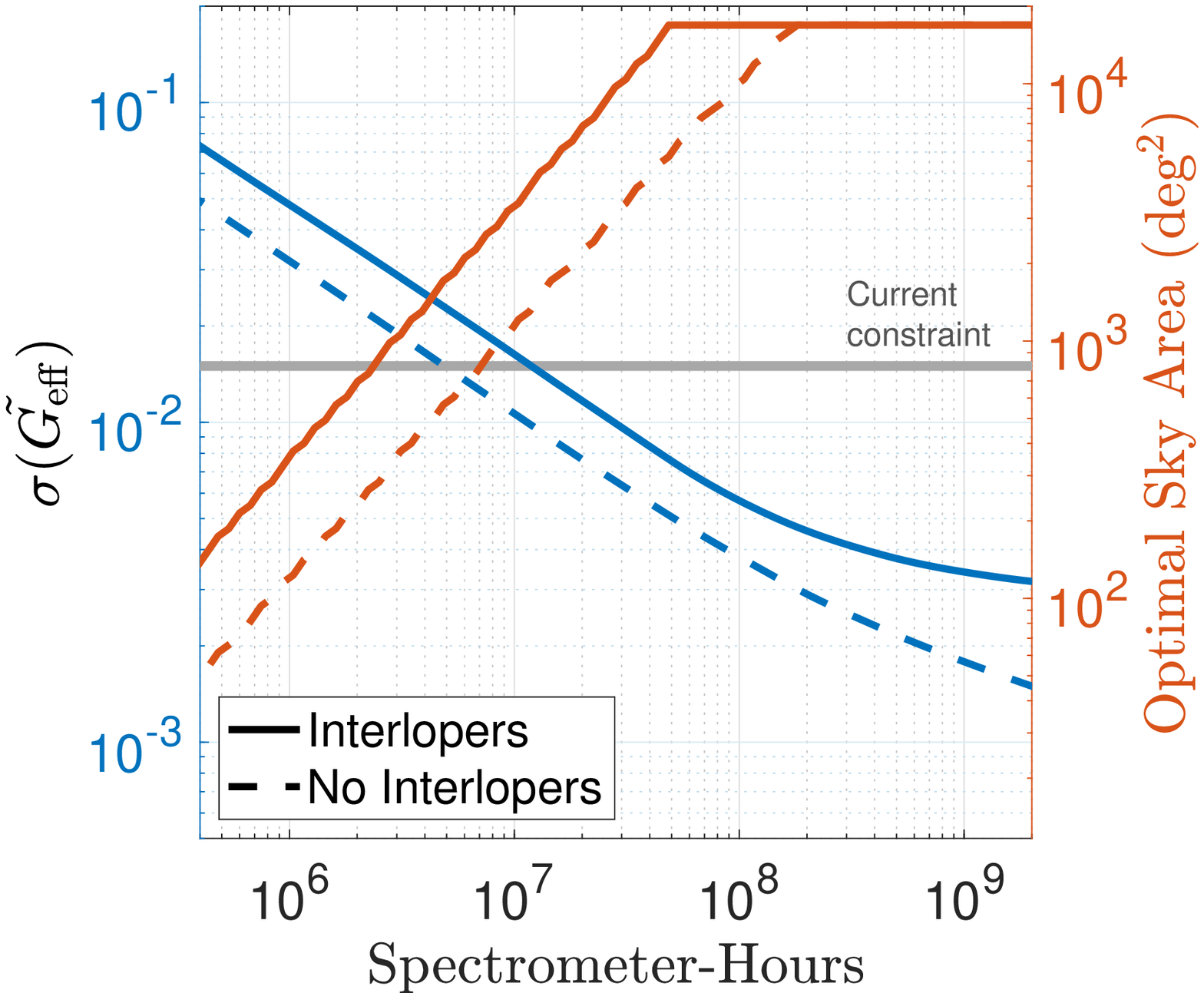}
        \caption{Blue lines show the 1-$\sigma$ constraints on JBD model parameters as a function of spectrometer hours, including (solid) and neglecting (dashed) interlopers. The red lines are the corresponding required sky coverage, and the grey bands are the existing constraints from a combination of KiDSx2dFLenS, BOSS full shape, Planck18 and Pantheon data \citep{Joudaki:2020shz}.}
    \label{fig:JBD_surveyopt}\vspace{0.1in}

    \centering
    \includegraphics[width=0.6 \textwidth]{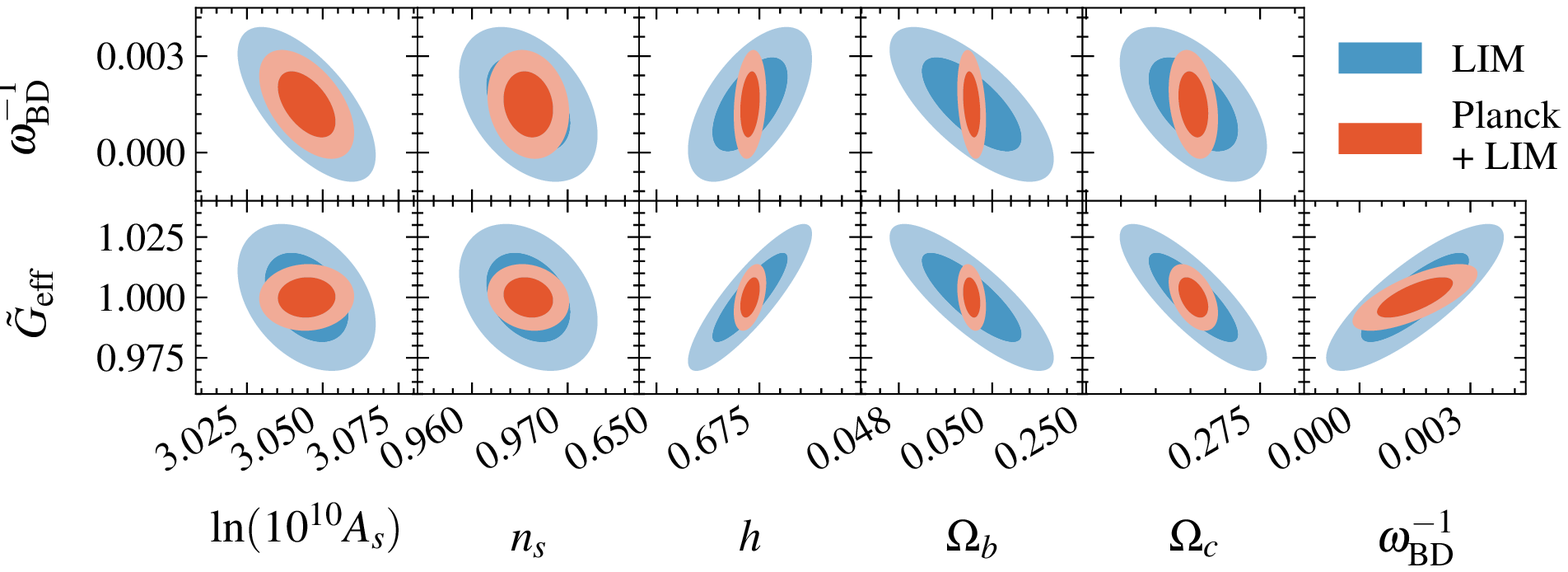}
    \vspace{-0.15in}
    \caption{Marginalized 1- and 2-$\sigma$ constraints on JBD and $\Lambda$CDM parameter pairs from LIM alone (blue) and when combined with $\Lambda$CDM Planck constraints (red). Interloper noise is included here, and the spectrometer hours is fixed to $\sim 10^8$.}
    \label{fig:jbd_2d}
    \vspace{0.1in}
\end{figure*}

In Fig.~\ref{fig:JBD_surveyopt}, we show the 1-$\sigma$ constraints on parameters of the JBD model as a function of spectrometer hours, assuming the fiducial value of $\omega_{\rm BD}^{\rm fid} = 800$ . The solid (dashed) blue lines show the constraints, including (neglecting) the interloper noise, and the red lines correspond to the required sky coverage to obtain those constraints. The plateau in the red lines corresponds to the maximum sky coverage of $f_{\rm sky} \simeq 0.5$ that we have imposed. The grey bands are the current constraint from \cite{Joudaki:2020shz}.

For the comparison with the current bounds, it should be kept in mind that the JBD constraints (in particular $\omega_{\rm BD}^{-1}$) depends on the choice of fiducial values. Our choice of $\omega_{\rm BD}^{\rm fid} = 800$ is comparable to the peak of the 1d posterior of $w_{\rm BD}^{-1}$ in \cite{Joudaki:2020shz}. While only upper bounds on $\omega_{\rm BD}$ are reported in \cite{Joudaki:2020shz}, the gray band we show here loosely corresponds to $\sigma(\omega_{\rm BD}^{-1}) \sim 0.002$. 

For $\lesssim 10^8$ spectrometer hours, the extra noise due to interlopers degrades the constraints on $\omega_{\rm BD}^{-1}$ and ${\tilde G}_{\rm eff}$ by less than a factor of 2. At higher values of spectrometer hours, not removing the interlopers causes saturation of the constraining power as the interloper noise dominates the thermal noise. Furthermore, the unsubtracted interloper noise increases the required sky coverage by nearly a factor of 2; without interlopers the maximum sky coverage of 50\% is saturated at $\sim 2\times 10^8$ spectrometer hours, while with the interloper noise, it is reached at $\sim 5 \times 10^7$.
Fig.~\ref{fig:jbd_2d} shows the 1- and 2-$\sigma$ marginalized constraints on parameter pairs to illustrate parameter degeneracies. Blue contours are from LIM only, while the red ones include Planck priors. As expected from the discussion in \S \ref{sec:th_MG_DE}, for LSS probes, there are strong degeneracies between ${\tilde G}_{\rm eff}, h$, and $\Omega_c$ (and to a lesser extent with $\Omega_b$) due to their similar effects on expansion rate and growth of structure.  Imposing Planck priors improves the LIM-only results, both by tightening the constraints on $\Lambda$CDM parameters and by alleviating parameter degeneracies. We caution that while here we imposed Planck $\Lambda$CDM priors, a more careful joint analysis of CMB and LIM, assuming the same model, is necessary to quantify the information content of the individual and combined probes more accurately.

\begin{figure}[htbp!]
    \centering
        \includegraphics[width=0.4 \textwidth]{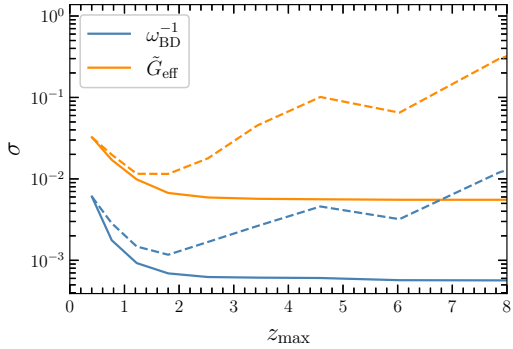}
        \vspace{-0.1in}
        \caption{Solid lines show the marginalized 1-$\sigma$ constraints on JBD model as a function of maximum redshift. For reference, we also show the constraints as a function of the center of redshift bins in dashed lines. Here, the interloper noise is accounted for, Planck priors are imposed, and the spectrometer hours is fixed to $\sim 10^8$.}
    \label{fig:jbd_sigma_z}
\end{figure}

Fig.~\ref{fig:jbd_sigma_z} shows the constraints as a function of maximum redshift $z_{\rm max}$ (top panel) and as a function of the center of each redshift bin (bottom panel), assuming a fiducial value of $\omega_{\rm BD} = 800$ and imposing Planck $\Lambda$CDM priors. For both parameters, the constraints improve with increasing $z_{\rm max}$ up to $z_{\rm max}\sim 2.5$, after which a near plateau is reached. At $z\sim 6$ when the [CII] signal also contributes, we see a decrease of $\sim 15\%$ on $\sigma(w_{\rm BD}^{-1})$, while $\sigma(\tilde G_{\rm eff})$ is only lowered by $3\%$. At high redshifts, the survey we have considered here only access to high-J rotational lines of CO, which are expected to be much fainter (in brightness temperature units) then their low-J counterparts. As a result, the constraints at $z\sim 6$ are primarily driven by [CII]. 

Comparing the constraints for fiducial values of $\omega_{\rm BD}^{\rm fid}=800$ and $\omega_{\rm BD}^{\rm fid}=10^5$, we find that the latter are tighter. To understand this behaviour, it is important to stress that the larger the value of $\omega_{\rm BD}$ is, the lesser the scalar field is evolving. In other words, in the limit $\omega_{\rm BD}\rightarrow\infty$ the scalar field is completely frozen and reduces to a cosmological constant. Therefore, requiring a value of $\tilde{G}_{\rm eff}$ different than unity today would imply almost the same value at earlier times. On the other hand, it is easier to accommodate deviations from $\tilde{G}_{\rm eff}=1$ for lower values of $\omega_{\rm BD}$, since an evolving scalar field would go back (in the past) to its tracker value ($\phi=1$) rapidly. This is reflected also in the constraints of the other parameters. As an intuitive example, one could roughly think that the constraints on the parameters of the $\omega_{\rm BD}^{\rm fid}=10^5$ model are the ones on the $\omega_{\rm BD}^{\rm fid}=800$ model, but fixing $\tilde{G}_{\rm eff}=1$, rather than marginalizing over it. Therefore, in comparing our results with previous forecasts, e.g., those of \cite{Alonso:2016suf}, the difference in the assumed fiducial values, should be kept in mind.

\begin{figure*}[htbp!]
    \centering
    \includegraphics[width=0.32\textwidth]{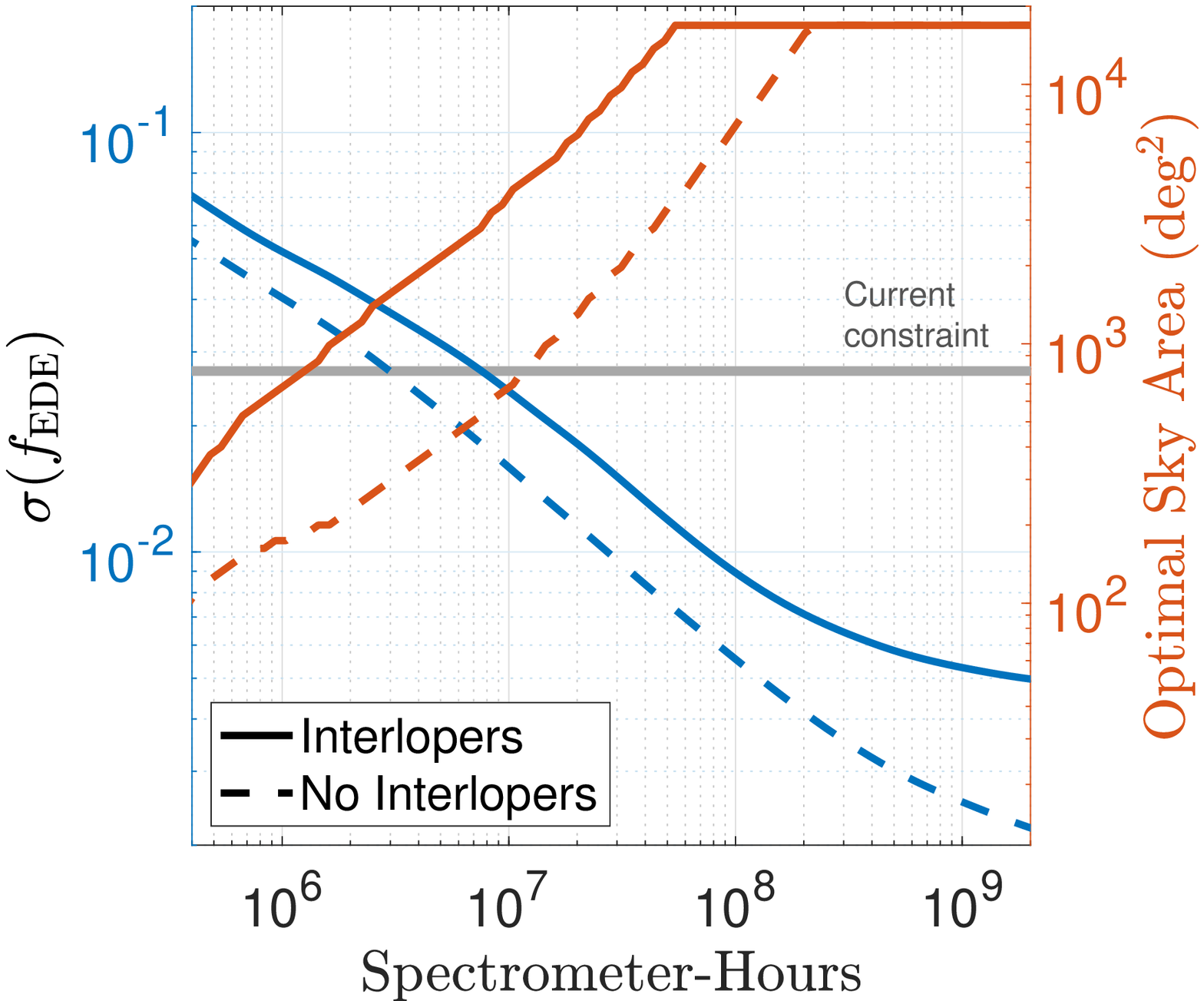}\vspace{.15in}
    \includegraphics[width=0.32 \textwidth]{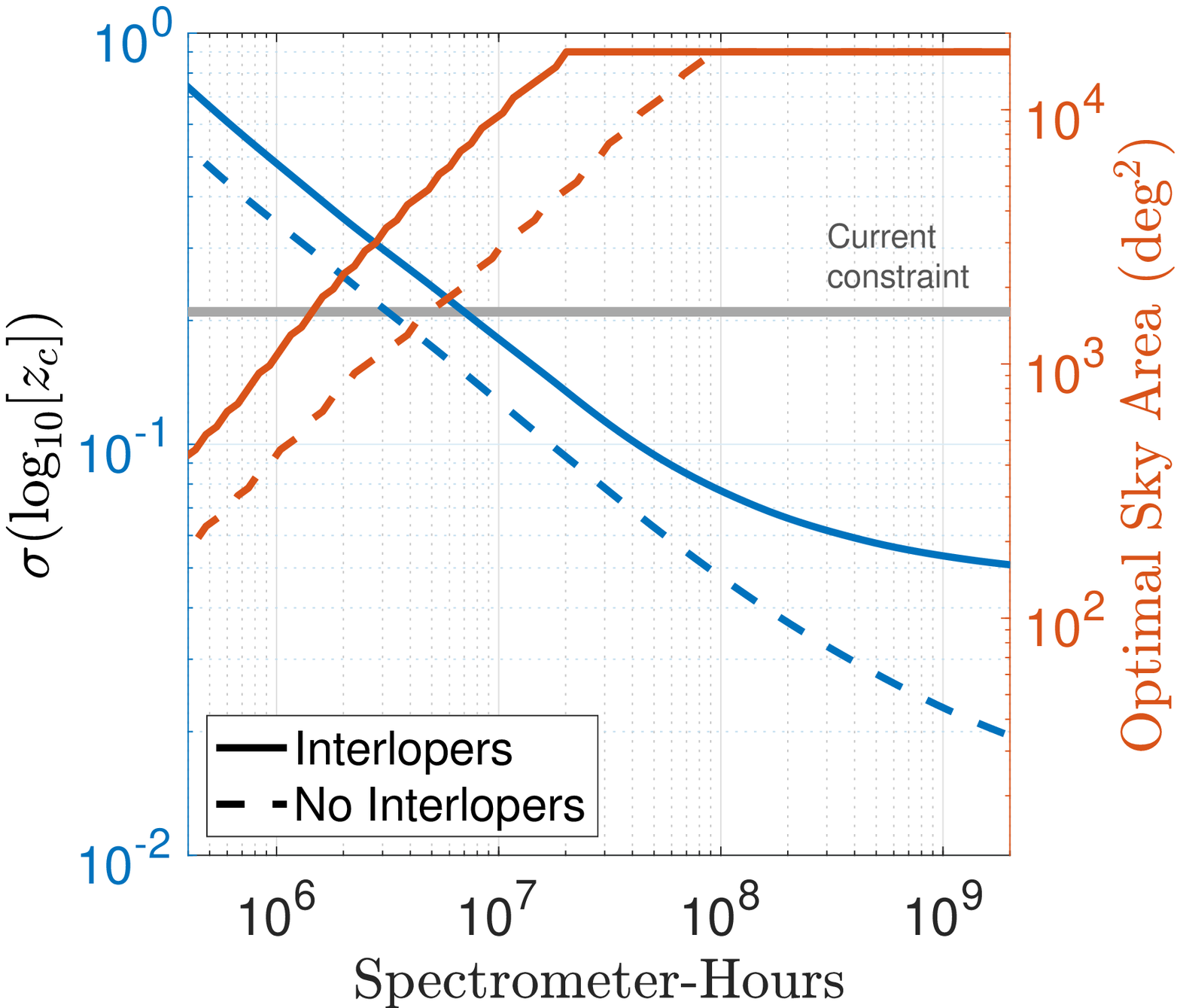}
    \includegraphics[width=0.32\textwidth]{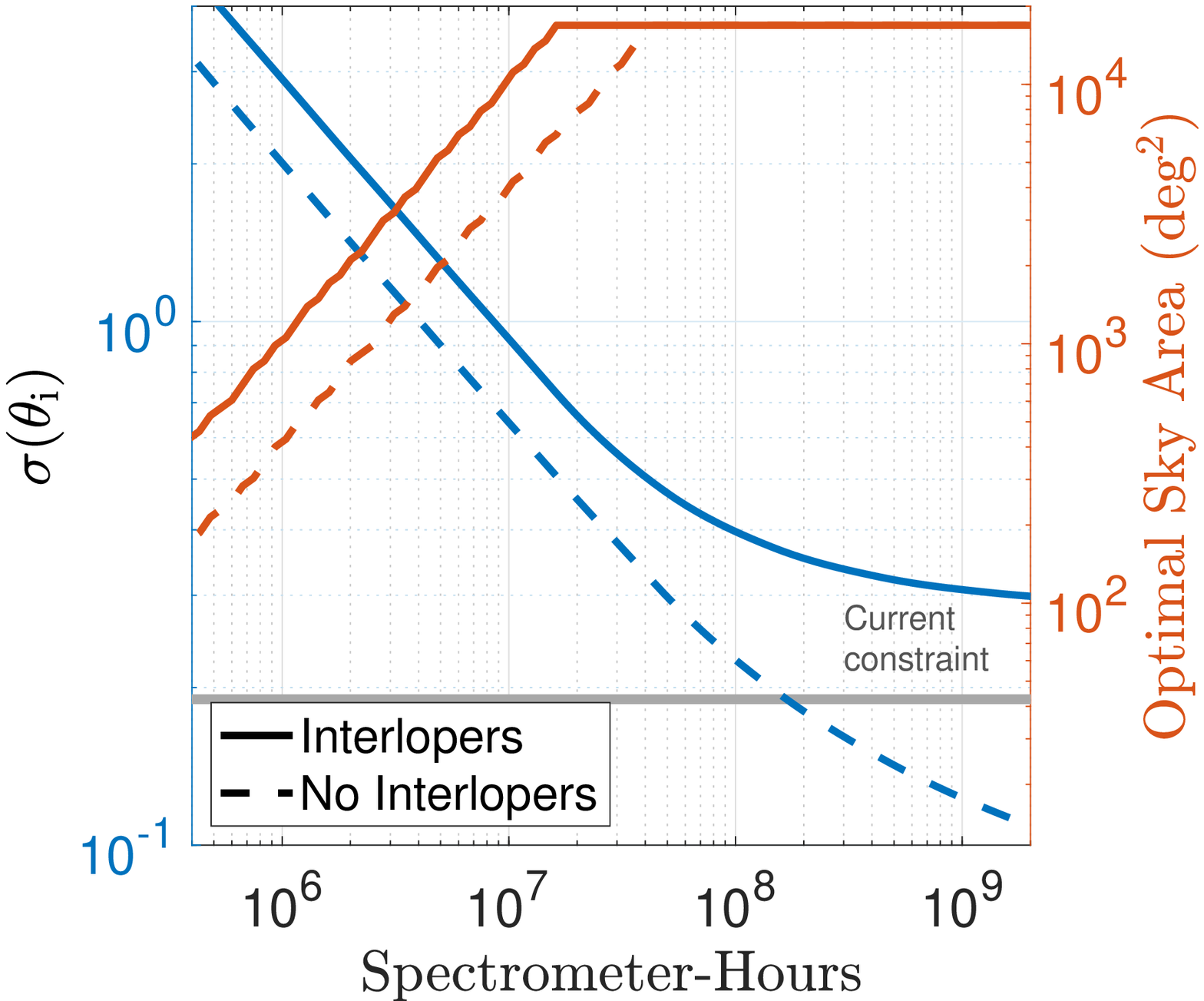}
    \vspace{-0.1in}
    \caption{Blue solid (dashed) lines show the marginalized 1-$\sigma$ marginalized constraints on EDE driven by ultra-light axions, including (neglecting) line interloper noise. The red lines show the corresponding required sky coverage, and the grey band shows the current bounds from Planck (temperature, polarization, and lensing) data and BOSS full-shape galaxy power spectrum, imposing prior on h0 from SH0ES local distance-ladder data \citep{Simon:2022adh}. The $\Lambda$CDM Planck priors are applied.} 
    \label{fig:EDE_surveropt}
\end{figure*}

\begin{figure*}[htbp!]
\centering 
\includegraphics[width=0.63\textwidth]{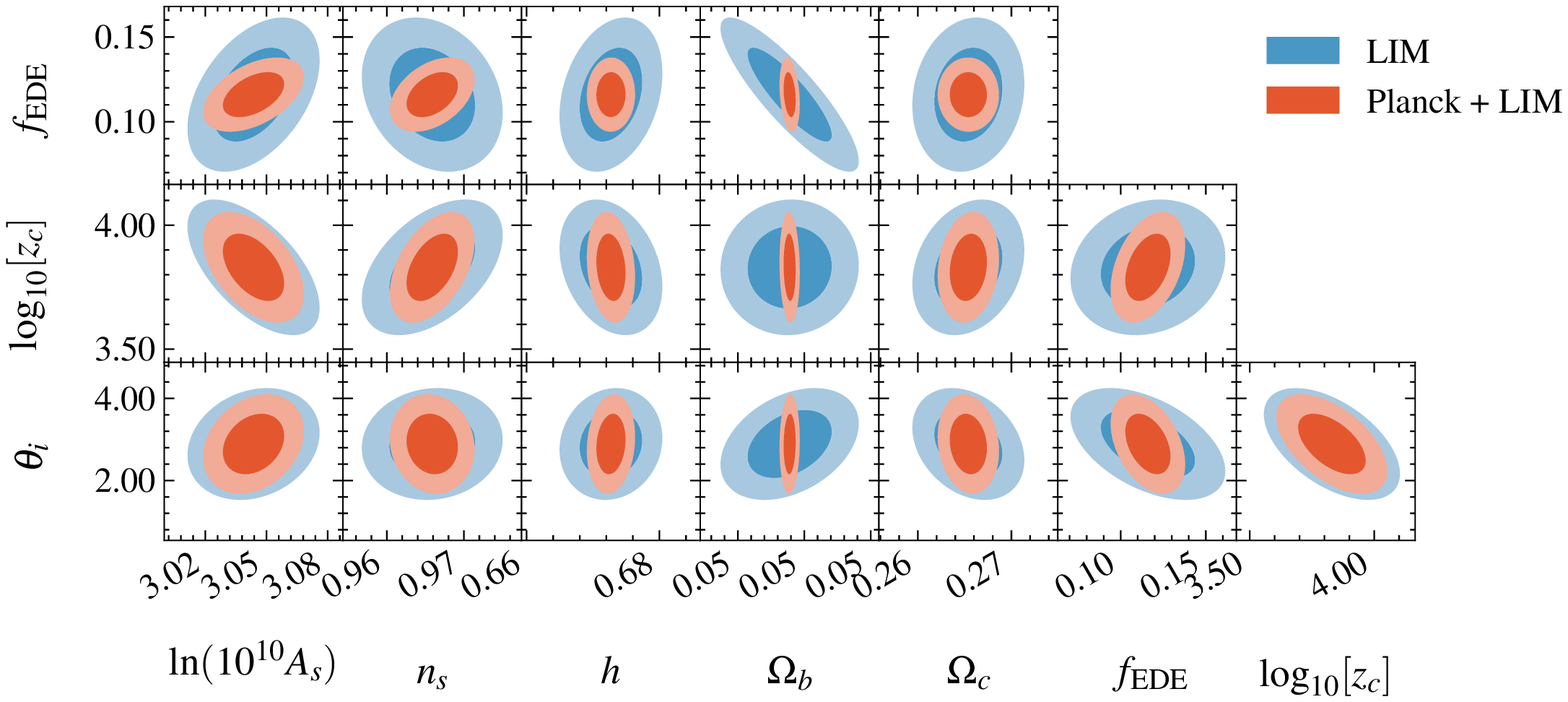}
\caption{Marginalized 1- and 2-$\sigma$ constraints on EDE and $\Lambda$CDM parameter pairs from LIM alone (blue) and when combined with $\Lambda$CDM Planck constraints (red). Interloper noise is included here, and the spectrometer hours is fixed to $\sim 10^8$.}
    \label{fig:ede_2d}\vspace{0.1in}
\end{figure*}

\begin{figure}[htbp!]
    \centering
        \includegraphics[width=0.38 \textwidth]{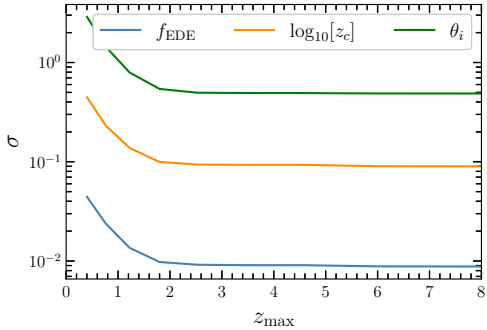}\vspace{-0.1in}
        \caption{Marginalized 1-$\sigma$ constraints on EDE model parameters as a function of maximum redshift, including interloper lines, imposing Planck priors, and fixing the spectrometer hours to $\sim 10^8$.}
    \label{fig:ede_sigma_z}
\end{figure}
\subsection{Early Dark Energy driven by ultra-light axions}

In Fig.~\ref{fig:EDE_surveropt}, we show the 1-$\sigma$ constraints on the three parameters of the EDE model as a function of spectrometer hours (in blue), and the corresponding sky coverage (in red). Again solid (dashed) lines show the results with (without) interloper noise, and the grey bands are the current constraints from \cite{Simon:2022adh}. As described in \S \ref{sec:ede}, in comparison of our results with existing bounds, caution should be taken since not having Planck constraints on EDE parameters and using only $\Lambda$CDM priors, makes our LIM+Planck constraints significantly more pessimistic. Keeping this in mind, we see that the constraints on $f_{\rm EDE}$ can be improved significantly over the current bounds. On the other hand, lowering the uncertainties for $\log_{10}(z_c)$ and in particular $\theta_i$ is more challenging, requiring large number of spectrometer hours and larger sky coverage. In addition to the point above about the comparison with the current bounds, one can argue that the poor constraint on $\theta_i$ is also driven by the fact that the signature left by a change in its value, i.e.\ the normalized initial value of the scalar field, is erased by the proceeding evolution, mainly due to the EDE feature at recombination. For $f_{\rm EDE}$ and $\log_{10}(z_c)$, removing the interloper noise would allow a factor of $\sim(4-5)$ improvement over the current bounds for $\sim 10^8$ spectrometer hour we considered, while the constraints on $\theta_i$ (without full Planck constraints) would be a factor of $\sim 2$ worse. 

In Fig.~\ref{fig:ede_2d}, we show the 1- and 2-$\sigma$ marginalized constraints on parameter pairs. There are several degeneracies between EDE and $\Lambda$CDM parameters in LIM-only results in addition to those among the EDE parameters. The most notable ones are between $f_{\rm EDE}$, and $\Omega_b, A_s$, and $n_s$, and between $\log_{10}(z_c)$, $A_s$, and $n_s$. Adding Planck priors on $\Lambda$CDM improves the constraints, in particular by tightening the constraints on $\Omega_b$.

\begin{figure*}[t!]
    \centering
    \includegraphics[width= 0.32\textwidth]{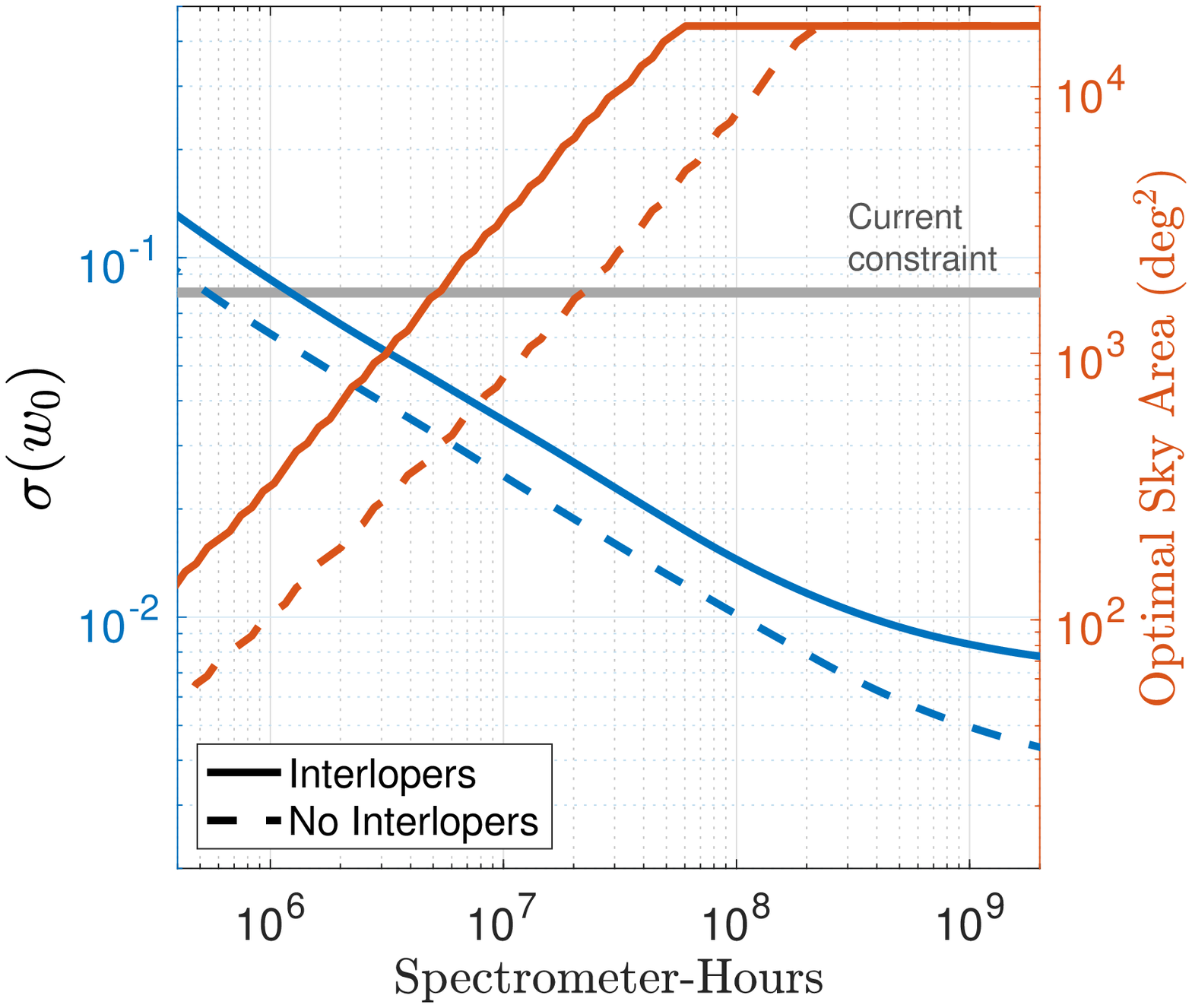}
    \includegraphics[width= 0.32\textwidth]{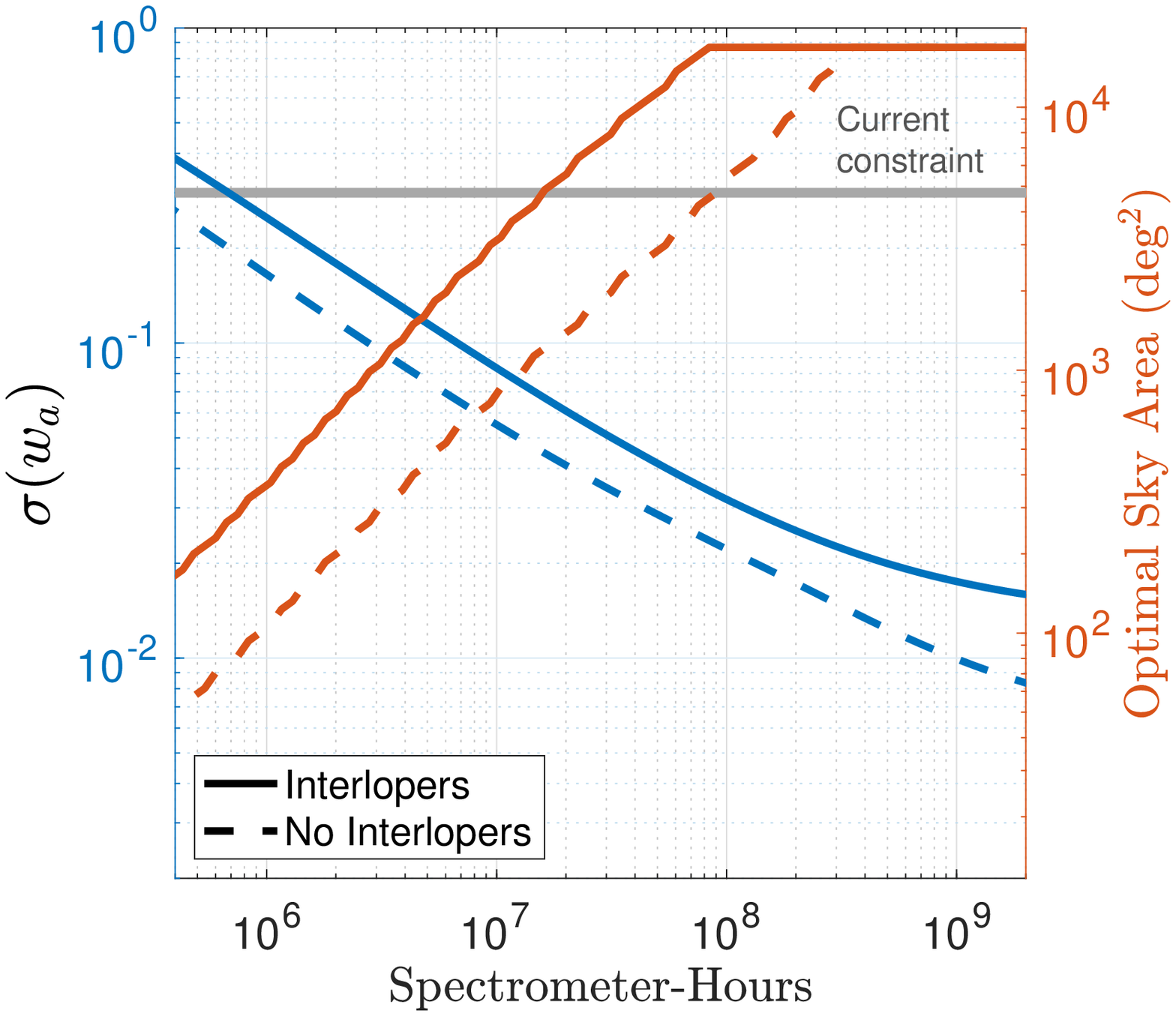}
     \caption{Blue solid (dashed) lines show the marginalized 1-$\sigma$ constrained and optimized survey configuration for CPL model, including (neglecting) interlopers, and imposing Planck priors on $\Lambda$CDM parameters. Red lines are the corresponding required sky coverage, and the grey bands are the current bounds from CMB, BAO, and Supernovae data.} 
     \label{fig:CPL_surveyopt}
\end{figure*} 
\begin{figure*}[htbp!]
    \centering
        \includegraphics[width= 0.6\textwidth]{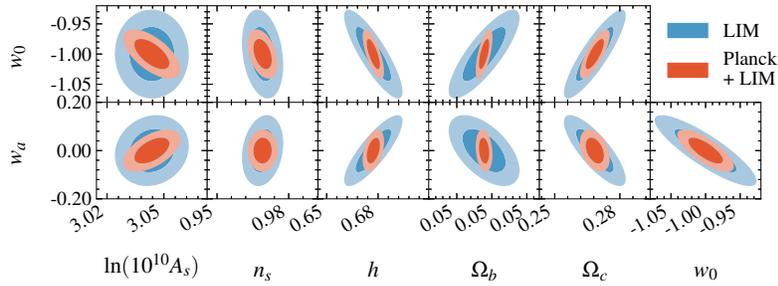}
        \caption{Marginalized 1- and 2-$\sigma$ constraints on CPL and $\Lambda$CDM parameter pairs from LIM alone (blue) and when imposing Planck priors on $\Lambda$CDM parameters. Interloper noise is included here. The sky fraction is assumed to be $f_{\rm sky} = 0.5$, and the spectrometer hours is fixed to $\sim 10^8$.}
    \label{fig:CPL_2d}\vspace{0.1in}
\end{figure*}
\begin{figure}[h]
    \centering
        \includegraphics[width=0.4 \textwidth]{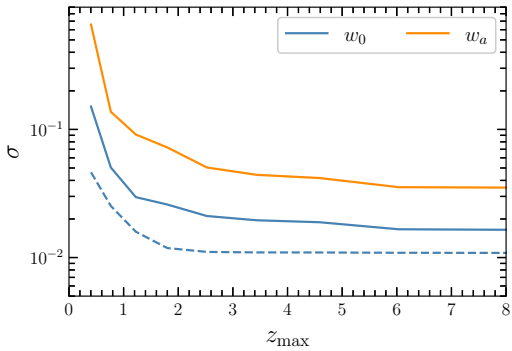}
        \caption{Marginalized 1-$\sigma$ constraints on CPL model as a function of maximum redshift, including interloper noise, imposing Planck priors, and fixing the spectrometer hours to $\sim 10^8$. The dashed blue line shows the constraint on $w_0$ assuming constant equation of state of DE.}
    \label{fig:CPL_sigma_z}
\end{figure}

In Fig.~\ref{fig:ede_sigma_z}, we show the 1-$\sigma$ constraints as a function of maximum redshift $z_{\rm max}$. Again, the trend is similar to that of the JBD model; constraints improve with increasing redshift up to $z_{\rm max}\lesssim 2.5$. After that, the constraints almost completely plateaus. Without imposing Planck priors (not shown in this plot), the constraint on $\log_{10}(z_c)$ improves by $\sim 14\%$ at $z_{\rm max}\sim 6$, at which [CII] signal is also included. Let us emphasize that in interpreting the information content of LIM measurements at different redshifts some care should be taken. First, we have only included the effects of EDE on liner matter power spectrum, and modeled the line power spectrum at linear level, and also did not include any modification to halo mass function due to and EDE \citep{Klypin:2020tud}. Second, for the survey specification that we have assumed, the thermal noise rapidly increases as we go to higher redshifts beyond $z\simeq 2.5$. Therefore, the fact that our results do not show a significant improvement beyond $z\simeq 2.5$ should not be taken as a strong statement about the constraining power of LIM measurements at high redshifts.

\begin{figure*}[!htbp]
    \centering
    \includegraphics[width=0.34 \textwidth]{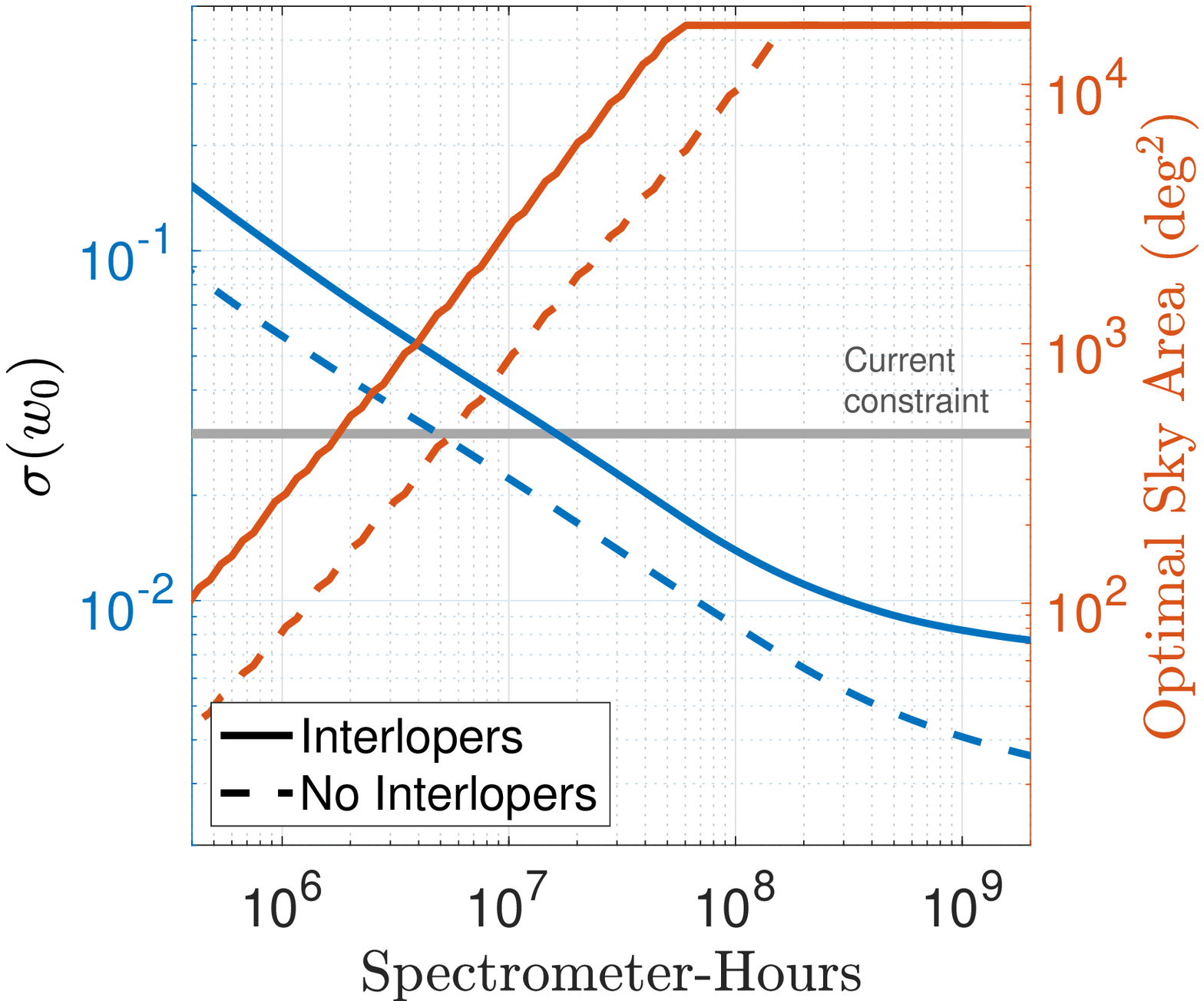}
    \includegraphics[width=0.34\textwidth]{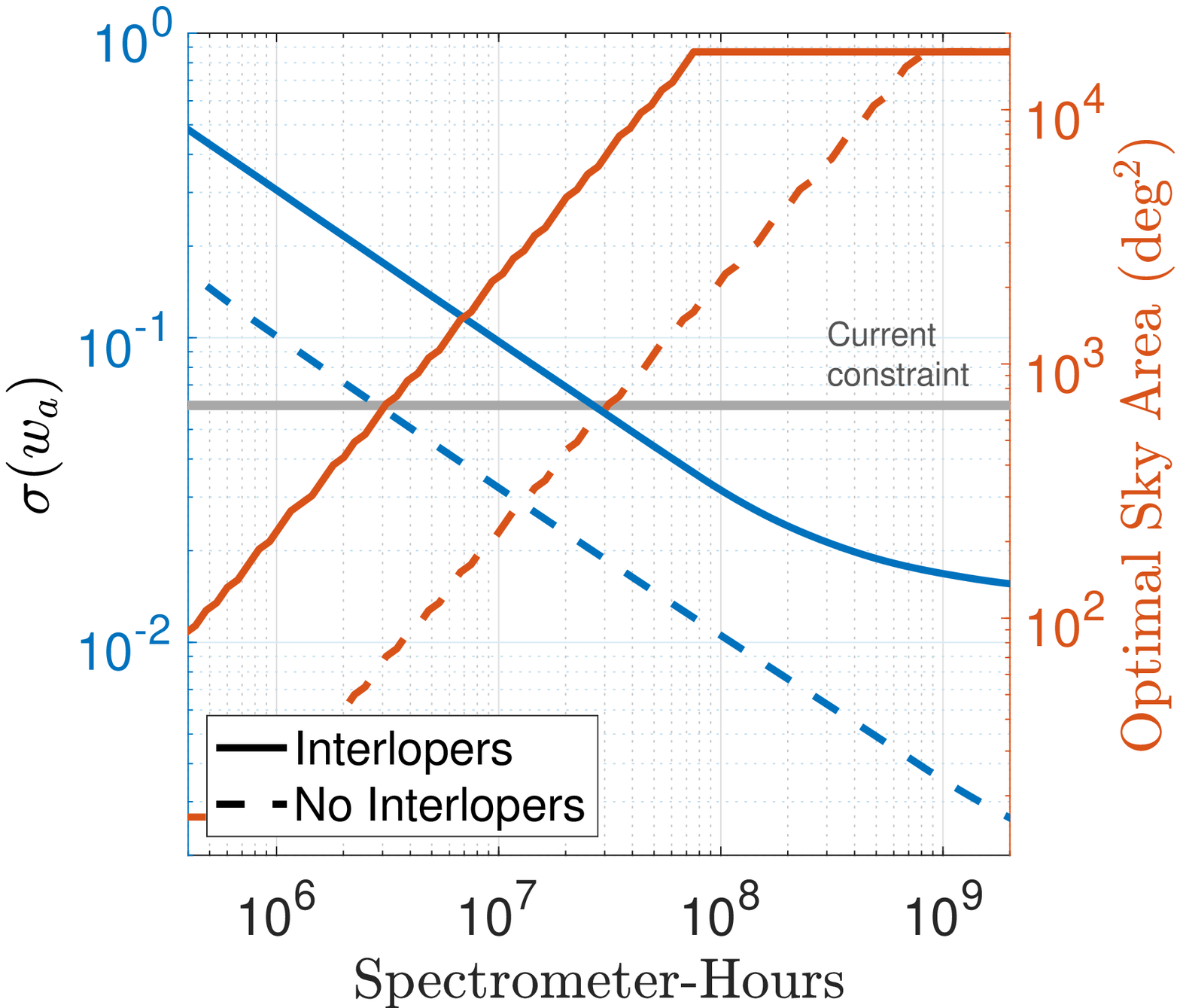}
    \includegraphics[width=0.34\textwidth]{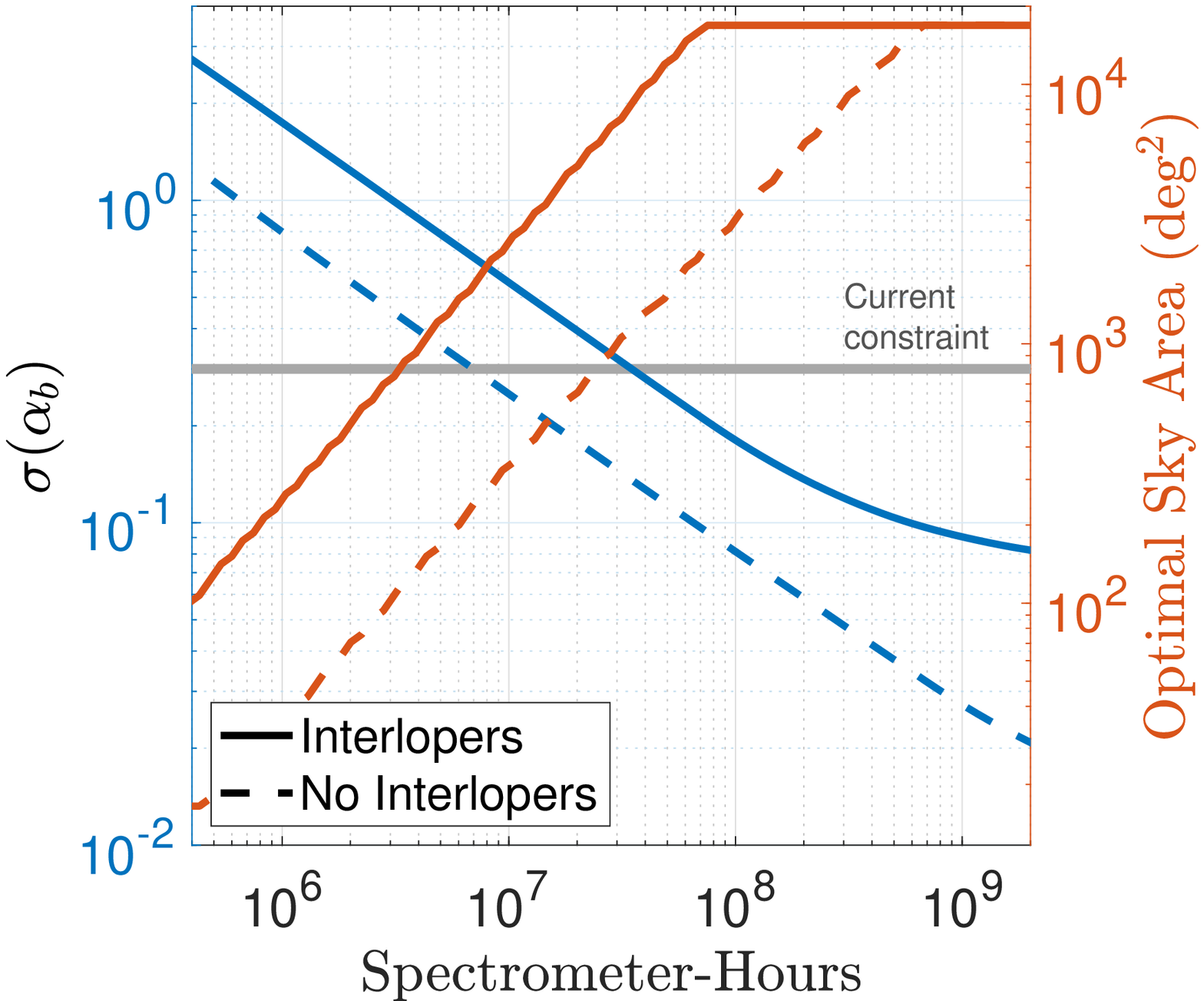}
    \includegraphics[width=0.34 \textwidth]{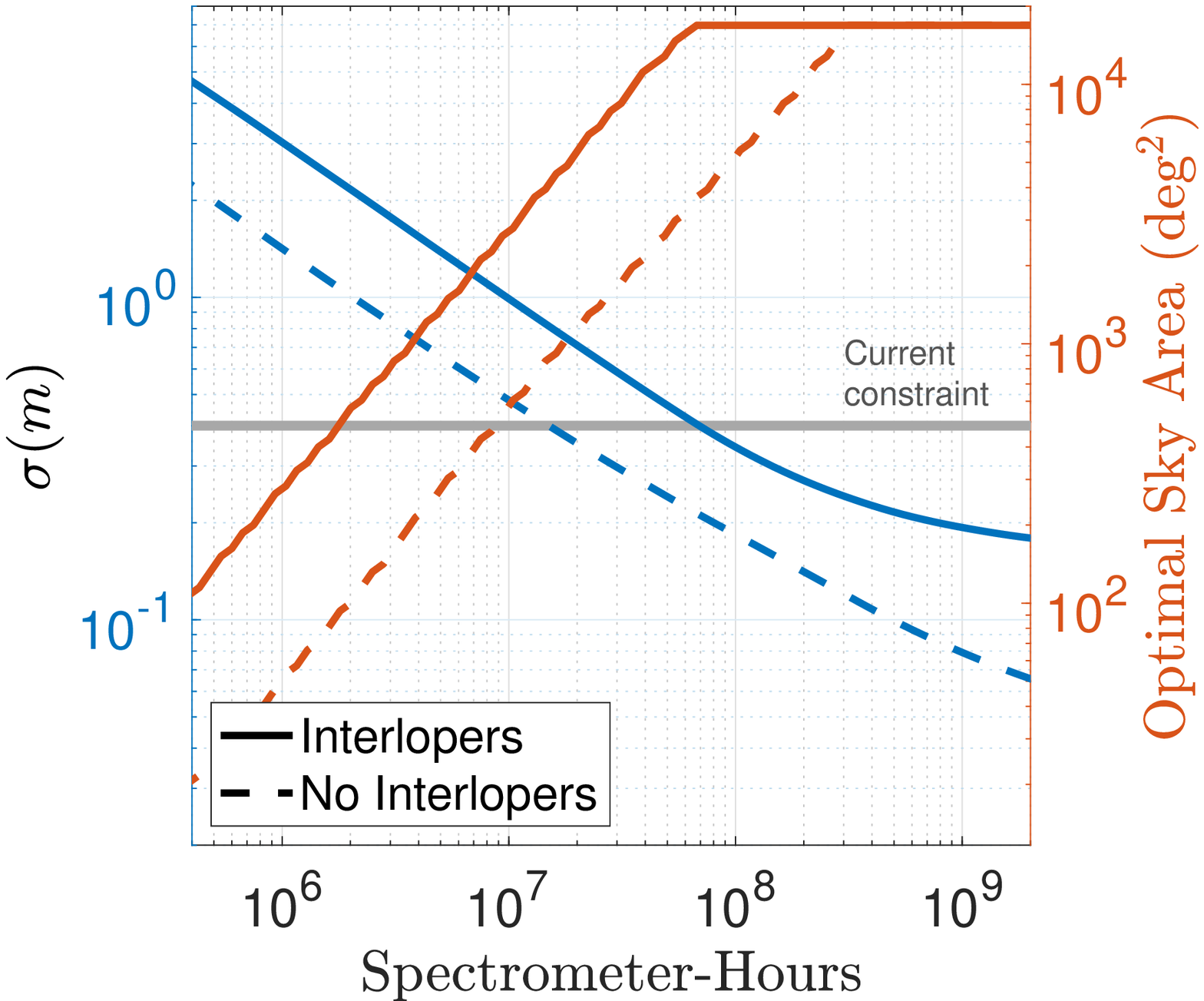}\
    \caption{Blue solid (dashed) lines show marginalized  1-$\sigma$ constraints on background (top) and perturbation (bottom) parameters of the shift-symmetric Horndeski model, including (neglecting) interloper noise. Planck priors on $\Lambda$CDM parameters are imposed. Red lines are the corresponding required sky coverage and grey bands are the current bounds from CMB, BAO, RSD and Supernovae Type IA data jointly with theoretical priors of \cite{Traykova:2021hbr}.} 
    \label{fig:ShiftSymm_surveyopt}
    \vspace{0.1in}
\end{figure*}

\subsection{Standard CPL parametrization of Dark Energy}\label{sec:cpl_results}

In Fig.~\ref{fig:CPL_surveyopt}, we show the results of survey optimization for the CPL model, imposing $\Lambda$CDM Planck priors and including (solid) or neglecting (dashed) interlopers. Similar trends to other models is observed, i.e.\ improvement of constraints as a function of spectrometer hours, and a near power-law increase of the required sky coverage. The considered survey can provide significantly tighter constraints in the $w_0-w_a$ plane than the current bounds from the combination of CMB, BAO, and Supernovae \citep{Planck:2018vyg}. For $\sim10^8$ spectrometer hours, and including the interloper noise, we obtain $\sigma(w_0) \sim 0.016$  and  $\sigma(w_a) = 0.034$. As shown in Fig.~\ref{fig:CPL_surveyopt}, even a relatively modest survey covering $\sim 10\%$ of the sky with $\sim 10^7$ spectrometer hours can improve the current bounds on $w_0$ and $w_a$ by factors of about 2 and 3, respectively. 

Fig.~\ref{fig:CPL_2d} shows the 2D marginalized constraints on CPL and $\Lambda$CDM parameters with (red) and without (blue) Planck priors. There are strong degenracies between $w_0$ and $w_a$, as well as between each of them and $h, \Omega_b$, and $\Omega_c$. Again here, imposing Planck priors on $\Lambda$CDM improves the constraints on CPL parameters, primarily by tightening $\Lambda$CDM constraints. In actual joint analysis of the two data sets varying the same model, it is expected breaking parameter degeneracies will also play essential role (see e.g., \citealt{MoradinezhadDizgah:2021upg}).

In Fig.~\ref{fig:CPL_sigma_z}, we show the 1-$\sigma$ constraints on $w_0$ and $w_a$ as a function of maximum redshift. We also show the constraint for the model with $w_a=0$ in dashed line to highlight the information gain with increasing $z_{\rm max}$ for the CPL model. While for a non-dynamic DE, there is no information gain at $z \gtrsim 1.8$, for the CPL model, the constraints keep on improving (only at $5-9 \%$ level in the redshift range of $2.5<z_{\rm max}< 6)$. The addition of [CII] signal at $z\sim 6$ improves constraints on both $w_0$ and $w_a$ by about $20\%$ and $25\%$, respectively.

Our results are comparable to the forecasts obtained for Euclid-like experiments. \cite{Euclid:2019clj} found that the 1-$\sigma$ errors on $w_0$ vary from 0.02 (optimistic scenario) to 0.04 (pessimistic scenario). This range of values improves current constraints and it is at the level of ours for a LIM experiment with $\sim 10^8$ spectrometer hours. The situation is a bit more favorable to LIM-experiments for $w_a$, where their quoted constraints are $0.1-0.2$. From the bottom panel of Fig.~\ref{fig:CPL_surveyopt} we notice that the next generation of LIM surveys improves easily these numbers, having a redshift excursion much higher than LSS surveys.

\begin{figure*}[t!]
    \centering
        \includegraphics[width= 0.65\textwidth]{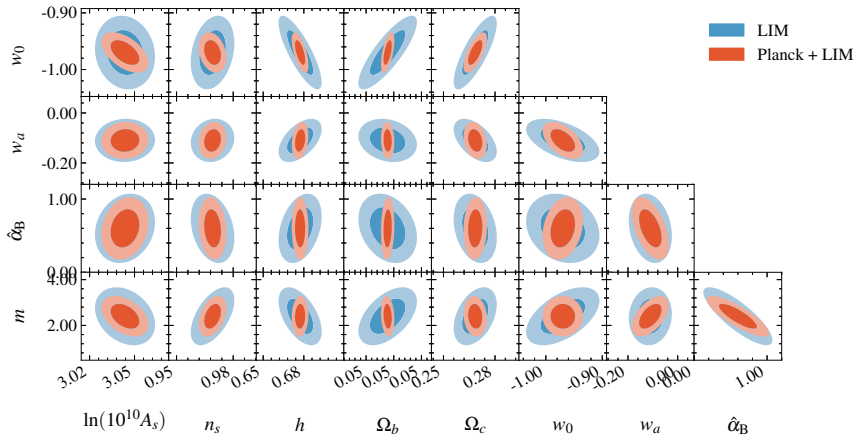}
        \caption{1- and 2-$\sigma$ marginalized constraints on ShiftSym-Horndeski vs. $\Lambda$CDM parameters from LIM alone (blue) and when imposing Planck priors on $\Lambda$CDM parameters. Interloper noise is included here. The sky fraction is assumed to be $f_{\rm sky} = 0.5$, and the spectrometer hours is fixed to $\sim 10^8$.}\vspace{0.1in}
    \label{fig:shiftsym_2d}
\end{figure*}
\begin{figure}[t]
    \centering
        \includegraphics[width=0.4 \textwidth]{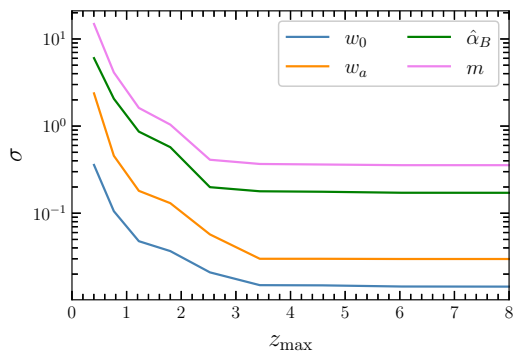}
        \caption{Marginalized 1-$\sigma$ constraints on ShiftSym-Horndeski model as a function of maximum redshift, including the interloper lines, imposing Planck priors, and fixing the spectrometer hours to $\sim 10^8$.}
    \label{fig:ShiftSym_sigma_z}
\end{figure}
\subsection{Shift-symmetric Horndeski Models}
In Fig.~\ref{fig:ShiftSymm_surveyopt}, we show the results of survey optimization for the shift-symmetric Horndeski model, imposing the Planck priors and including (solid) or neglecting (dashed) interlopers. Comparing with the constraints on $w_0$ in this model w.r.t.\ the corresponding one for the pure CPL parametrization (shown in the top plot of Fig.~\ref{fig:CPL_surveyopt}), we can see that the expected constraints with interlopers (solid blue lines) are essentially the same. This indicates that opening up the parameter space with $\hat{\alpha}_{\rm B}$ and $m$ does not degrades the constraining power of LIM experiments for the effective constant equation of state, as one would naively expect. This clearly suggests that the new parameter have orthogonal effects when considering LIM observables. It is also important to stress that what we quote here as ``current constraint'' for $w_0$ is lower than the corresponding constrain for the CPL model in \S \ref{sec:cpl_results}. This is simply due to the fact that in obtaining these constraints different datasets are used. 
Similar observation can also be made for $w_a$, with the only remarkable difference that at low spectrometer hours its constraints here are slightly worse than in the previous analysis. This suggests that the constraining power of current LIM surveys leaves some degeneracy that broaden the allowed parameter space. Considering $\hat{\alpha}_{\rm B}$, LIM alone will be able to improve the current constraints for $\gtrsim 4\times10^8$ spectrometer hours. With the parameter $m$ the situation is a bit different. First, the minimum spectrometer hours needed to improve the current constraints are $7\times10^8$; second, there is an almost-plateau at high spectrometer hours, which signals a diminishing return in pushing the survey specifications to their limits in constraining $m$.

Fig.~\ref{fig:shiftsym_2d} shows the 1- and 2-$\sigma$ marginalized constraints on the shift-symmetric Horndeski and $|Lambda$CDM parameter pairs, fixing the number of spectrometer hours to $\sim 10^8$. The LIM-only (blue) contours display strong degeneracies between $w_0$ and $h$, $\Omega_b$ and $\Omega_c$ as in Fig.~\ref{fig:CPL_2d}. Also, $w_a$ has some mild to large degeneracy with the same parameters. The degeneracy between $w_0$ and $\left(\hat{\alpha}_{\rm B},\,m\right)$ is removed when adding Planck priors, which tightly constrain $h$ and $\Omega_b$. The constraints on $w_0$ do not degrade with respect to CPL because the additional parameters are only mildly degenerate with $w_0$ when imposing Planck priors. Finally, it is worth noticing a strong degeneracy between $\hat{\alpha}_{\rm B}$ and $m$, which is due to the fact that the time modulation parameterized by $m$ can be reabsorbed into the overall amplitude $\hat{\alpha}_{\rm B}$ for small redshift variations.

Fig.~\ref{fig:ShiftSym_sigma_z} shows the marginalized $1\sigma$ constraints as a function of redshift. The improvements of the constraints for dynamical-DE parameters, $w_0$ and $w_a$ is significant up to $z\simeq 3.5$. Higher redshifts have little to none constraining power, similarly to the other models. It is worth noticing that here the constraints saturate at higher redshifts w.r.t.\ those of Fig.~\ref{fig:CPL_sigma_z}. This is because we chose a different fiducial model. Instead of a pure $\Lambda$ behaviour, here we have an evolving DE component, which leaves clear imprints at higher redshifts. The saturation of the constraining power for $\hat{\alpha}_{\rm B}$ and $m$ is reached at lower, but still large, redshift ($z\simeq2.5$). It is easy to show that the time evolution of $\alpha_{\rm B}\left(t\right)$ goes to zero more slowly than the most commonly considered models of modified gravity/DE which are designed to have low redshift signatures.

\subsection{Effective Description of \\ Luminal Horndeski Models}

\begin{figure*}[htbp!]   
    \centering
    \includegraphics[width=0.32\textwidth]{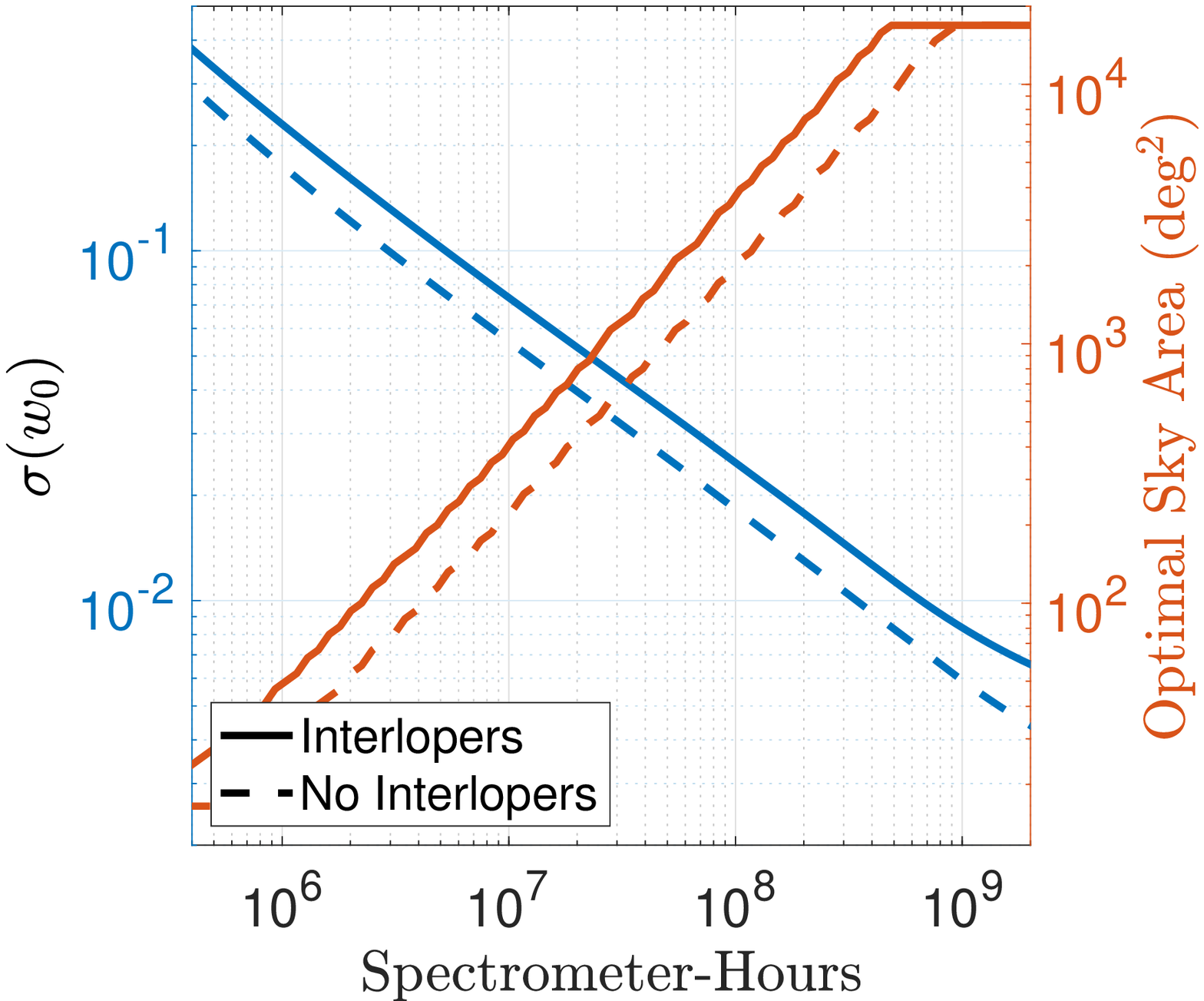}
    \includegraphics[width=0.32\textwidth]{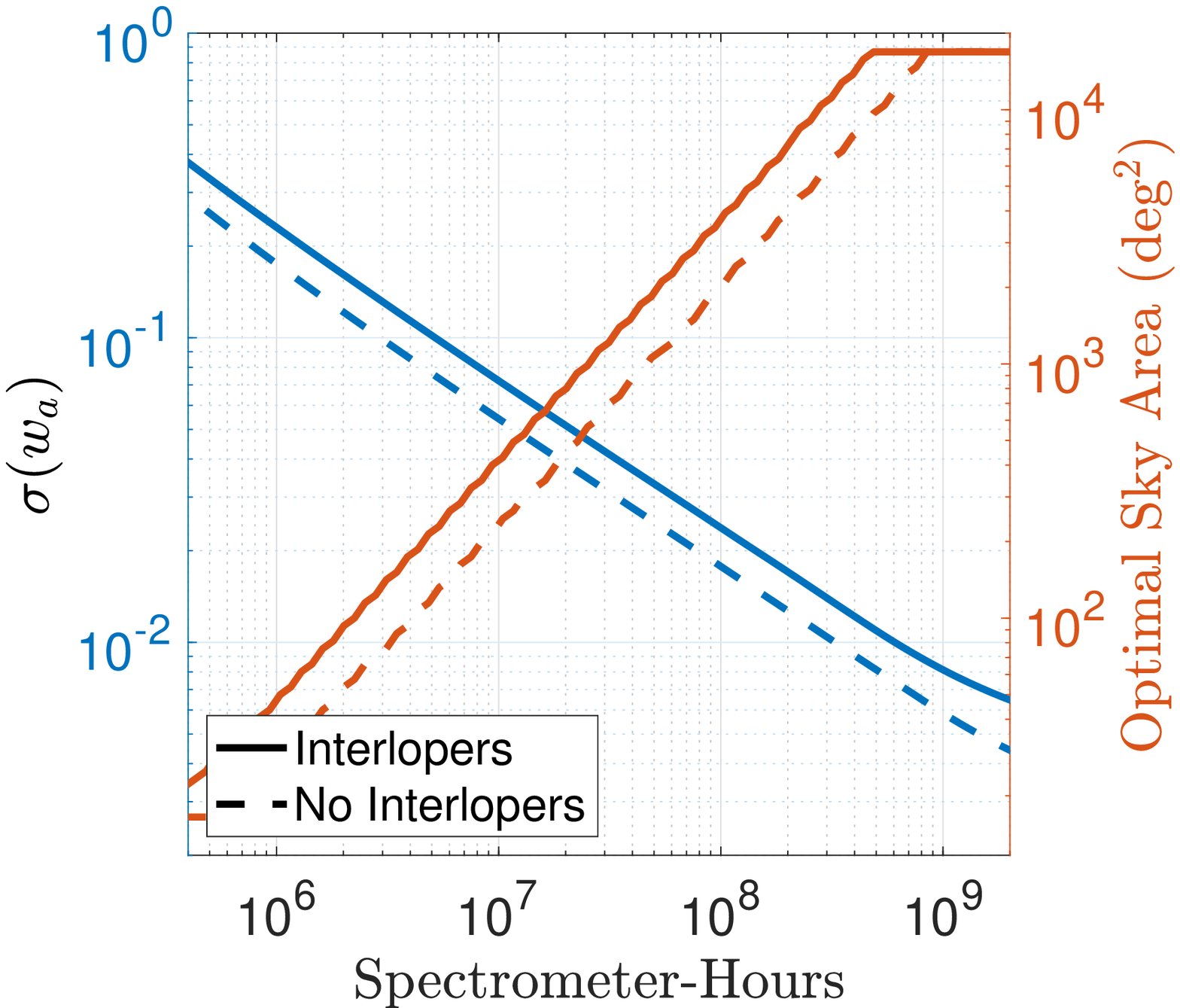}\\
    \includegraphics[width=0.32\textwidth]{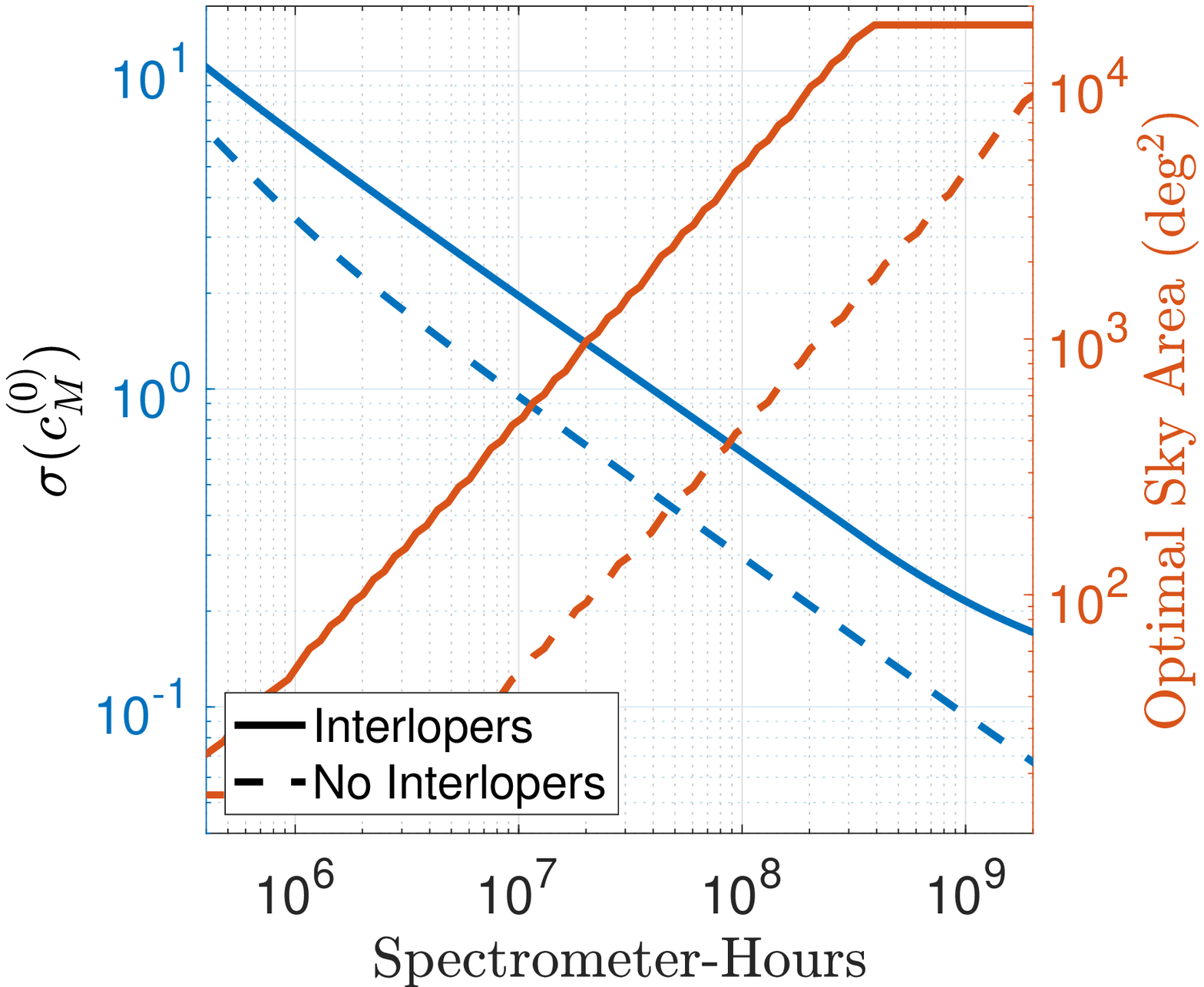}
    \includegraphics[width=0.32 \textwidth]{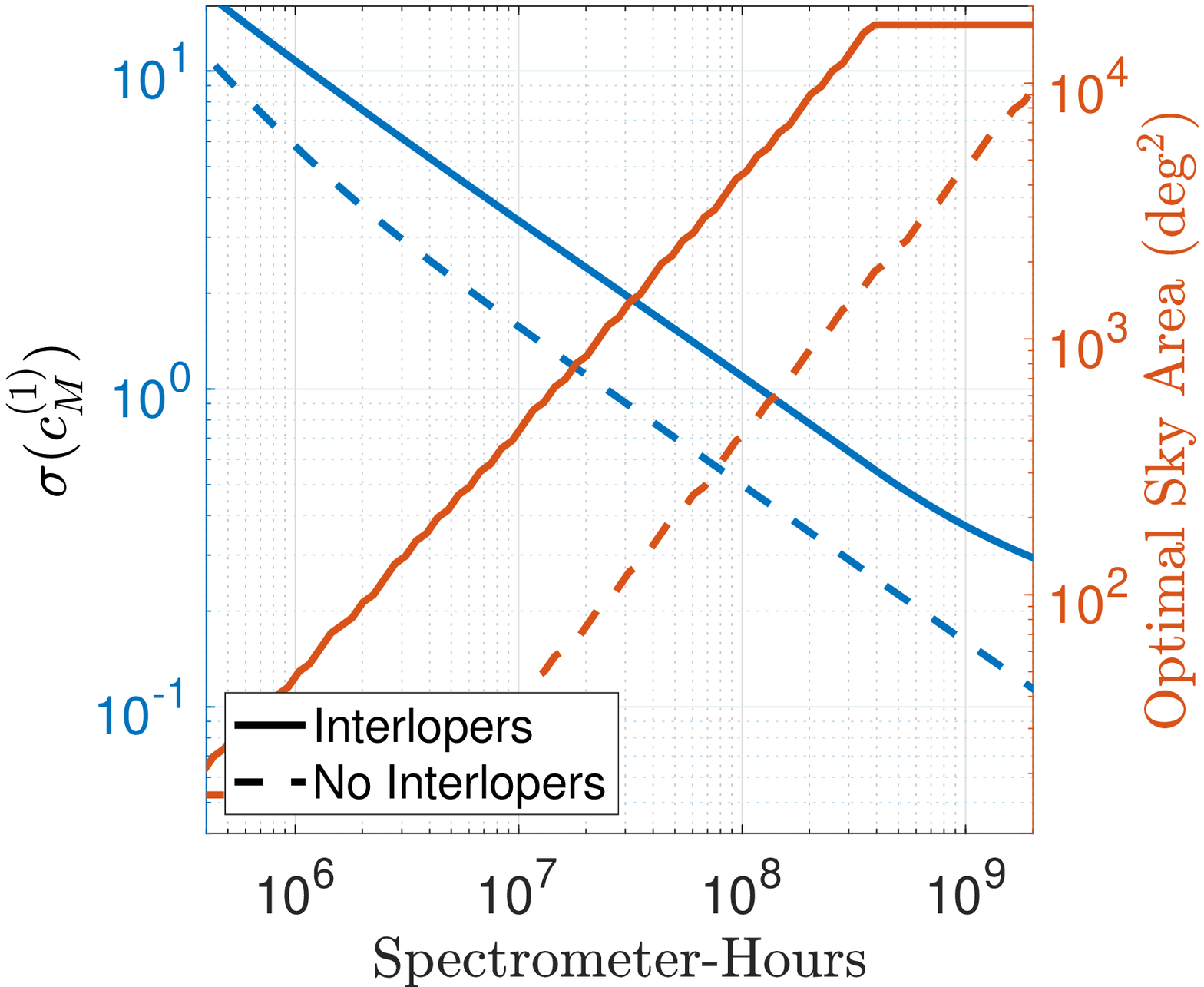}
    \includegraphics[width=0.32 \textwidth]{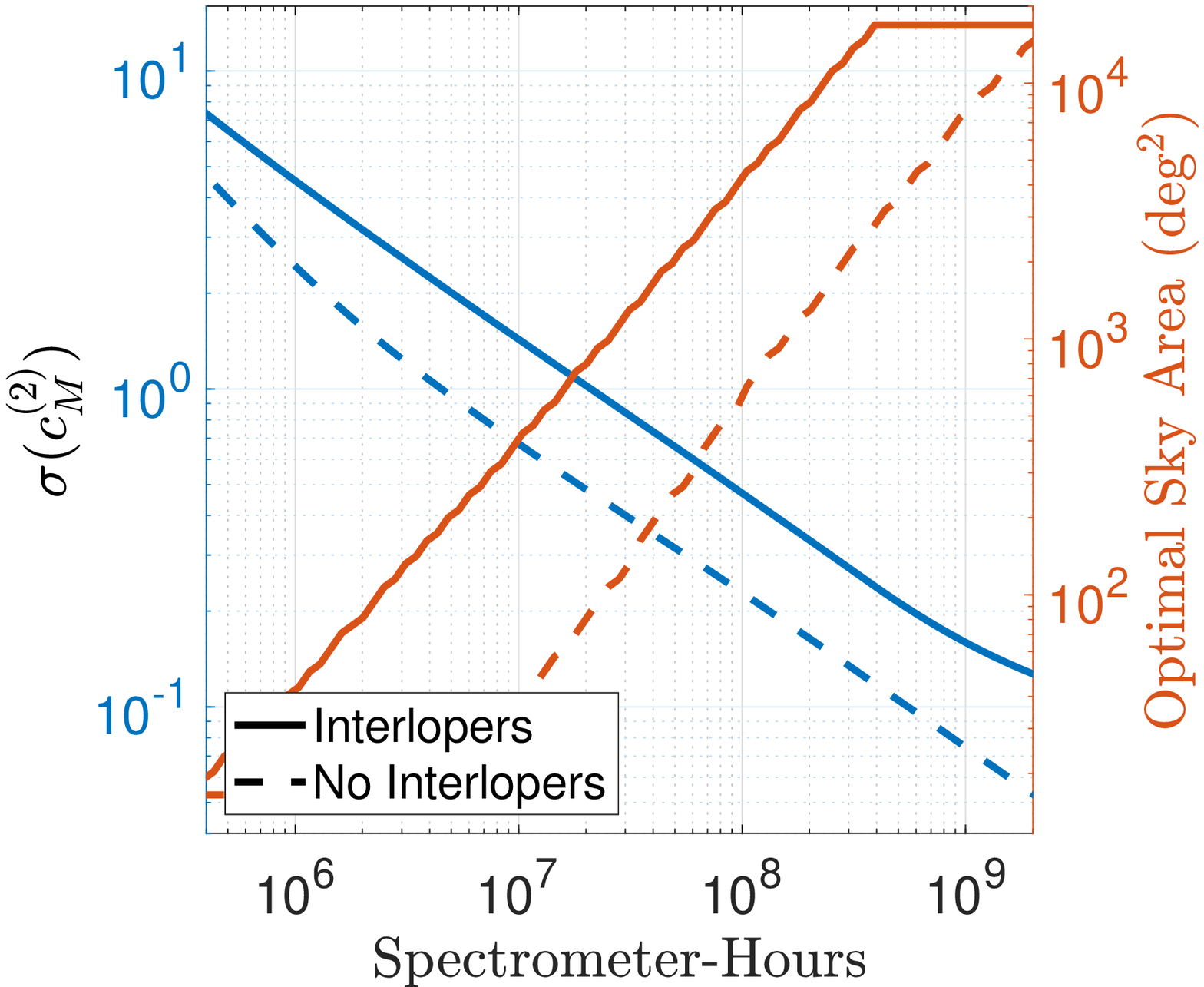}
    \caption{Blue solid (dashed) lines show marginalized 1-$\sigma$ constraints on background (top row) and perturbation (bottom row) parameters of the Effective-DE {\it model1}, including (neglecting) interloper noise. Red lines are the corresponding required sky coverage. Planck $\Lambda$CDM priors are imposed.}
    \label{fig:effDE1_BG_surveyopt}
    \vspace{0.1in}
\end{figure*}

\begin{figure*}[htbp!]
    \centering
    \includegraphics[width= 0.7\textwidth]{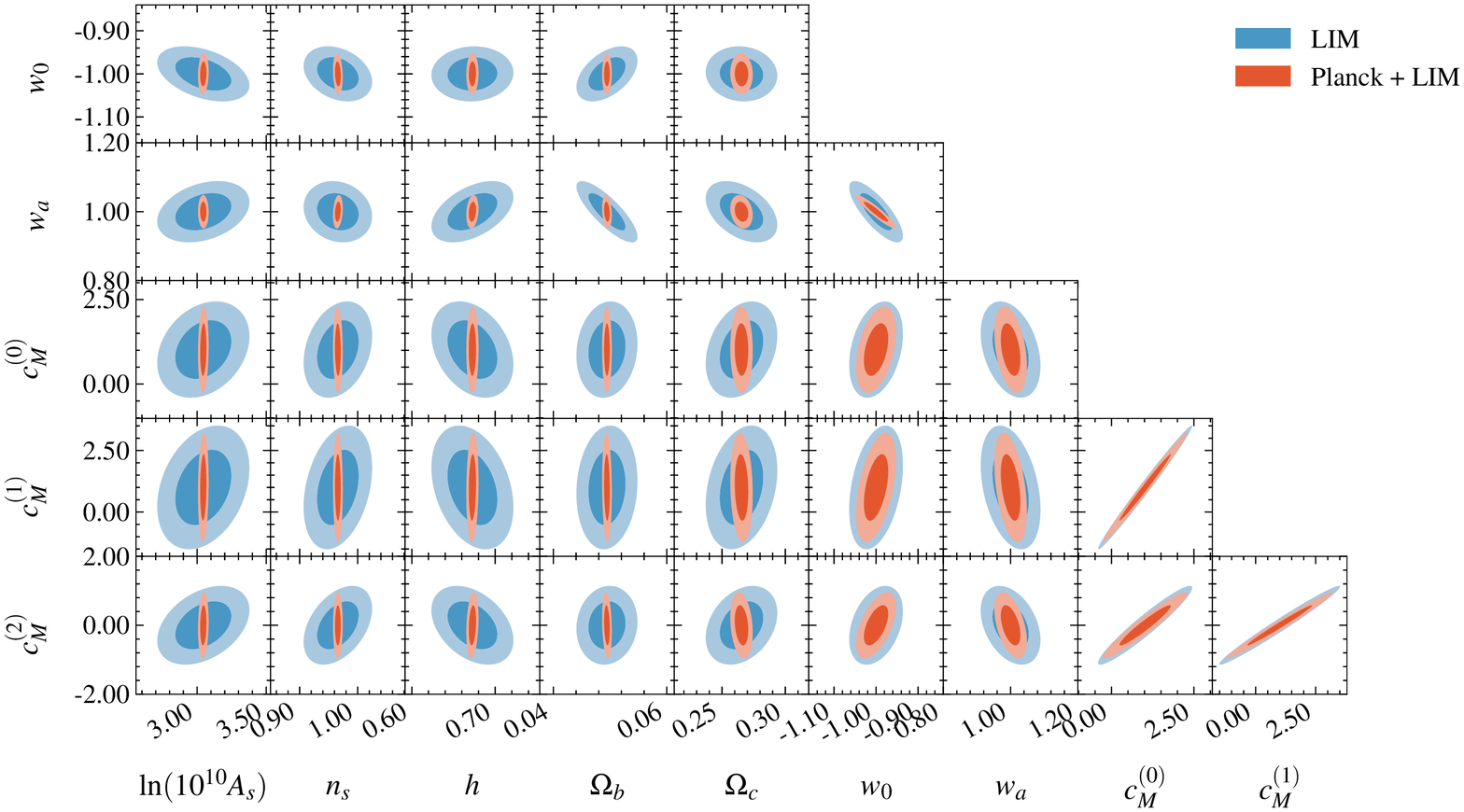} 
    \includegraphics[width= 0.7\textwidth]{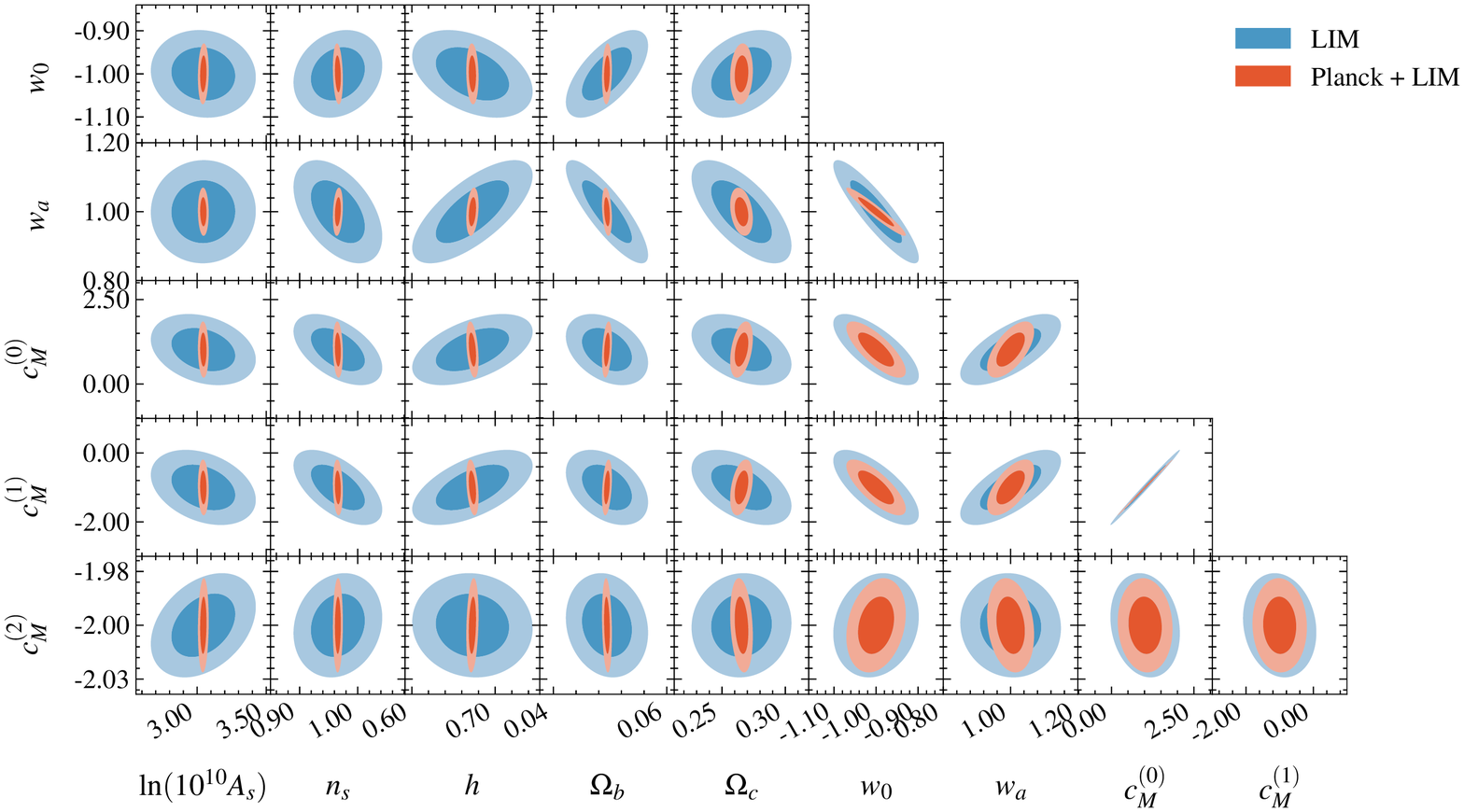}
    \caption{Marginalized 1- and 2-$\sigma$ constraints on  Effective-DE models and $\Lambda$CDM parameter pairs with (red) and without (blue) Planck $\Lambda$CDM priors. The top plot shows the constraints for {\it model1} while the bottom one correspond to {\it model2}. Interloper noise is included here, and the spectrometer hours is fixed to $\sim 10^8$.}
    \label{fig:eftDE_2d}
\end{figure*}

\begin{figure*}[htbp!]
    \centering
        \includegraphics[width=0.4 \textwidth]{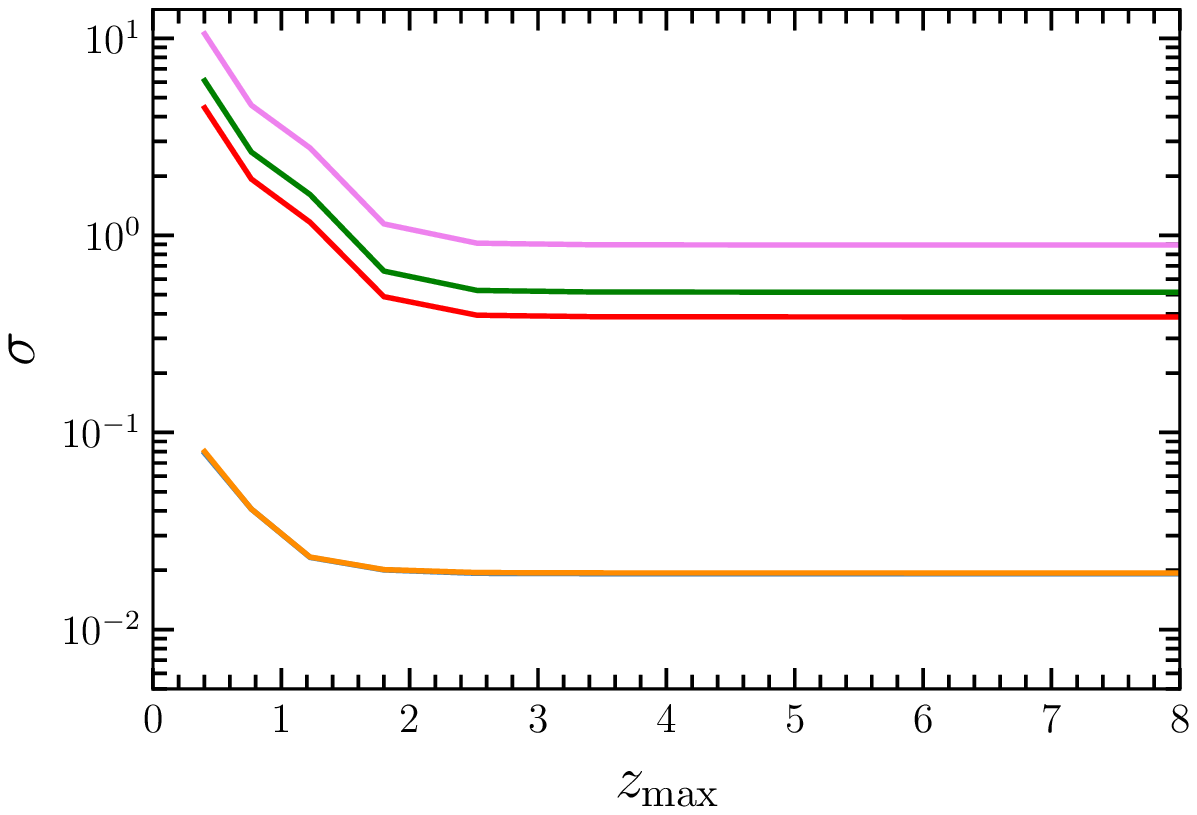}
        \includegraphics[width=0.4 \textwidth]{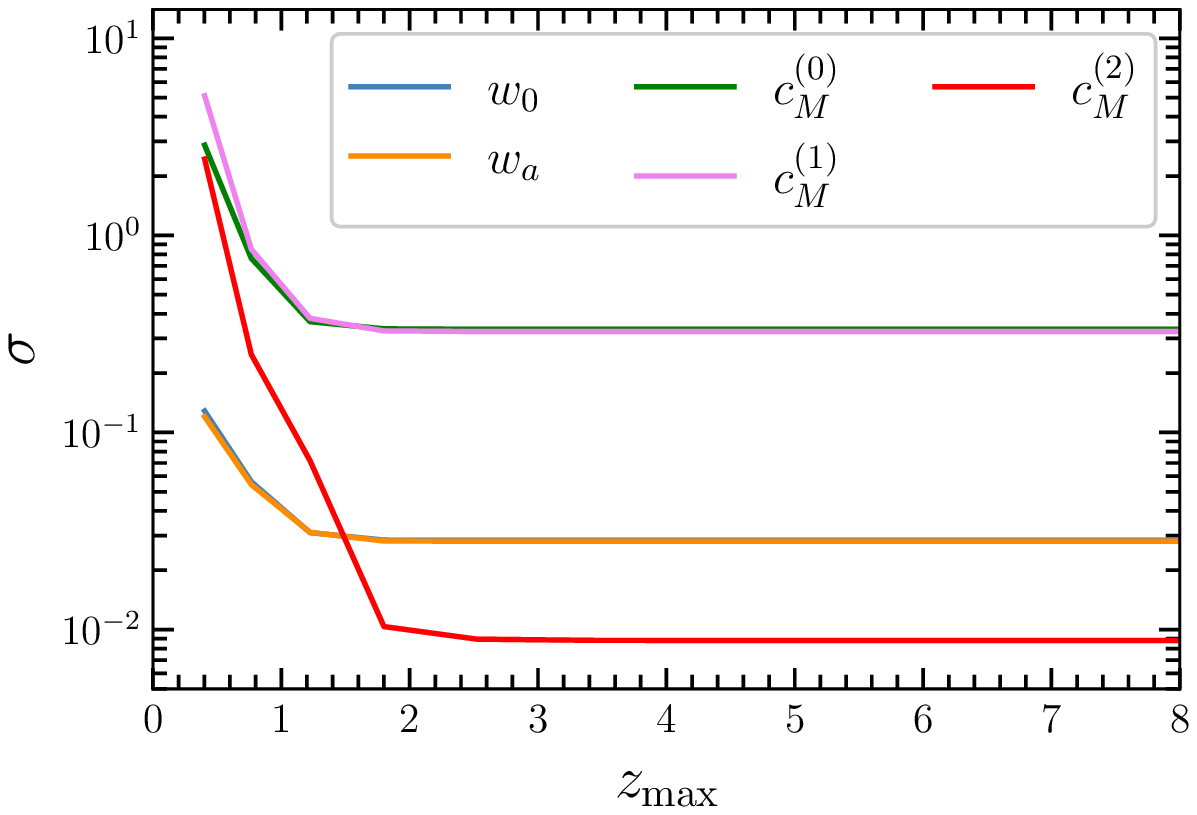}
        \caption{Marginalized 1-$\sigma$ constraints on Effective-DE {\it model1} (left) and {\it model2} (right) as a function of maximum redshift, including interloper noise, and imposing Planck priors. The curves for $w_0$ and $w_a$ (in both plots) and curves for $c_M^{(0)}$ and $c_M^{(1)}$ (in the right plot) lie nearly completely on top of each other. Here, interloper noise is included, Planck $\Lambda$CDM priors are imposed, and the spectrometer hours is fixed to $\sim 10^8$.}
    \label{fig:eftDE_sigma_z}
\end{figure*}

The effective description of DE is, to some extent (as discussed in \S\ref{sec:th_MG_DE}) the most general class of models we are investigating. On top of that, our fiducial models were designed \textit{ad-hoc} to have signatures beyond $z\simeq1$ as normal dynamical-DE models. It is important to stress that the analysis of these models is meant to be illustrative of the potential of LIM surveys to probe general EFTofDE models, rather than providing constraints on a specific model. Given our chosen parametrization, shown in Eqs. \eqref{eq:EFTofDE},  constraints on this class of models provide insights into evolution of dark energy equation of state, as well as time evolution of Planck mass. As described in \S \ref{sec:th_MG_DE}, both considered models are designed to have signatures at high-redshift, i.e., $2 \lesssim z \lesssim 10$. In particular, the effective Planck mass of {\it model1} increases smoothly with time, deviating from unity at the level of  $\sim 10\%$ at $z \simeq 10$. {\it model2} is even more extreme, with deviations $\sim 25\%$ at $z \simeq 10$, but also a feature at $z \simeq 1$ after which Planck mass starts decreasing. 

Fig.~\ref{fig:effDE1_BG_surveyopt} shows the results of survey optimization; the marginalized 1-$\sigma$ constraints as a function of spectrometer hours, together with the required sky coverage \footnote{While the plots refer to {\it model1}, most of the considerations that we are going to make are valid for {\it model2} as well. Whenever there are substantial differences between the results of the two models we will make it clear in the text.}. For all parameters, similar to all other models considered, the constraints improve as we increase the spectrometer hours. However, contrary to previous models, here removing the interlopers seems to be less important and we do not see saturation of the constraints in the range of spectrometer hours considered. This suggests that it should be possible (even if unrealistic with the next generation of surveys) to extract more information on these parameter increasing the observation time. One notable difference between {\it model1} and {\it model2} (not shown here) is the role of the interloper lines; perfect removal of the interlopers does not  considerably improve the constraints in {\it model2}, while it does for {\it model1}. 

In Fig.~\ref{fig:eftDE_2d}, we show the 1- and 2-$\sigma$ marginalized constraints for {\it model1} (top panel) and {\it model2} (bottom panel). Without Planck priors, the degeneracies between background ($w_0$ or $w_a$) and perturbation parameters ($c_M$ coefficients ) are less severe in {\it model1} than in {\it model2}. We note that in {\it model1}, the $c_M$ coefficients are highly correlated with one another, while in  {\it model2}, only the degeneracy between $c_M^{(1)}$ and $c_M^{(0)}$ is significant. The degeneracy directions between $c_M^{(0)}$ and $c_M^{(1)}$ coefficients and $w_0$ and $w_a$ in the two models are opposite one another. This is due the choices of fiducial values of the $c_M$ coefficients in the two models. The constraint on $c_M^{(2)}$ in {\it model1} is significantly weaker than in {\it model2} due to its sever degeneracies with $c_M^{(0)}$ and $c_M^{(1)}$. 

Fig.~\ref{fig:eftDE_sigma_z} shows the marginalized 1-$\sigma$ constraints as a function of redshift for {\it model1} (left) and {\it model2} (right). While the perturbation parameters are better constrained in {\it model2}, the constraints on background parameters are tighter in {\it model1}. The redshift dependencies of the constraints on $w_0$ and $w_a$ are similar in both model, mostly saturating at $z\simeq1.5$. For {\it model2} (right), the constraints on $c_{\rm M}^{(2)}$ improve sharply (and by $\sim$2 orders of magnitude) as a function of redshift. To interpret this result, one should take into account that the fiducial value for $c_{\rm M}^{(2)}$ in {\it model1} was zero, while in {\it model1} we chose a value different from zero. This makes small variations around the fiducial model much more important in the second model.

\section{Conclusion}\label{sec:conc}

Elucidating the nature of the dark energy component driving the current accelerated expansion and searching for possible modifications of gravity are two of major science goals for upcoming optical galaxy surveys. By offering the possibility of probing the LSS over a wide range of redshifts and scales, line intensity mapping can provide significant improvements in our understanding of the theory of gravity and the component of the Universe's energy density acting as DE, in  the regime largely beyond the reach of stage IV galaxy surveys. 

In this paper, we performed Fisher forecasts to determine the potential of a hypothetical future multi-stage ground-based LIM survey, targeting several rotational lines of CO (from J=2-1 to J=6-5) and the fine structure line of [CII], in constraining several beyond $\Lambda$CDM models. The considered models include two classes; the first class are consistently described by a specific covariant Lagrangian for a scalar field such as Jordan-Brans-Dicke gravity and (pre-recombination) early DE model. The second class are described in terms of effective description of the evolution of the background and perturbations, without defining a Lagrangian. An hybrid approach has been used for shift-symmetric Horndeski models. i.e., instead of constraining the parameters appearing in the Lagrangian, we used the corresponding effective description. For each model, we varied two survey specifications, the number of spectrometer hours and the fraction of sky covered by the survey, and optimized the survey strategy to obtain the best parameter constraints minimizing the survey cost. We considered optimistic and pessimistic scenarios for removal of line interlopers vs. accounting for them as a source of (anisotropic) noise, and obtained the parameter constraints with and without imposing Planck priors on $\Lambda$CDM parameters. 

Our results show that for all models, the considered LIM survey can improve upon the current bound by a factor of a few to an order of magnitude. The parameter uncertainties decrease as one increase the number of spectrometer hours and required sky coverage grows nearly linearly with decreasing 1-$\sigma$ errors. The interloper noise increases the required sky coverage to compensate for additional noise contribution. Including the interloper lines degrade the constraints on all models, but affects some parameters more significantly than others, causing a saturation of the parameter constraints even when increasing the number of spectrometer hours. Imposing Planck priors, thus tightening constraints on $\Lambda$CDM parameters, improves on  beyond $\Lambda$CDM parameters, especially those most degenerate with them.  

Among the models considered, for the JBD, the axion-driven EDE, and the CPL models, the constraints improve with increasing $z_{\rm max}$ up to $z_{\rm max} = 2.5$, after which a near plateau is hit. Addition of [CII] signal at $z\simeq6$ can improve the constraints on some of the parameters by at most (15-25)\%. For shift-symmetric Horndeski model, the constraints on DE equation of state keeps on improving up to higher redshifts ($z_{\rm max}\simeq 3.5$) and plateaus afterwards. For our selected effective description of luminal Horndeski models, the impact of interlopers doesn't lead to saturation of the constraining power at high spectrometer hours, and instead just degrade the constraint by nearly constant factors. The interlopers affect the sky required sky coverage for best constraints on parameters describing the time evolution of Planck mass more significantly than the dark energy equation of state parameters. 

It is worth noting that for several science goals considered in \S \ref{sec:res}, a millimeter-wave LIM experiment begins to produce constraints that are competitive with current measurements at $10^{7}$ spectrometer-hours, concurrent with previous work focused on LIM-enabled constraints on neutrino properties \citep{MoradinezhadDizgah:2021upg}. With a dedicated multi-year survey (with total observing time of $\gtrsim10^{4}$ hours), such a survey could be completed with a broadband camera of several dozen pixels -- only a few times larger than LIM-focused instruments currently being deployed \citep[e.g.][]{Karkare:2021ryi}. As demonstrated, such surveys would optimally cover a survey area of a few thousand square degrees, well within the reach of current survey telescopes (e.g., SPT).

There are several directions in which this work can be extended. First, here we have assumed that the line bias and mean brightness temperature are perfectly predicted by the theory. In reality, given the uncertainty on the theoretical predictions of these parameters (e.g. due to uncertainty in the model of line luminosity), one should marginalize over them. Second, since the variation of bias and mean brightness temperature is expected to degrade the constraints due to their degeneracies with cosmological parameters, studying how much the cross-correlations between different emission line signals, as well as between LIM and galaxy clustering and the CMB lensing can ameliorate these degeneracies is of great interest. On a more general ground, quantifying the potential improvements of the inferred constraints resulting from including both auto and cross-correlations is highly relevant. Third, in order to to go beyond the Fisher forecasts, and in the absence of current wide-field LIM survey, performing likelihood analysis of mock LIM data, imposing Planck priors on the specific beyond $\Lambda$CDM models considered is necessary. While here we have used Planck priors on $\Lambda$CDM parameters, joint analysis of LIM and CMB primary anisotropies assuming the same model for the two is essential. Fourth, we have assumed a linear model of the line power spectrum, and limited our analyses to relatively large scales. Including nonlinearities and promoting the model to include one-loop corrections, and marginalizing over a complete set of line biases is a natural next step. The improved model will allow us to go to smaller scales, and potentially help in breaking the degeneracies between mean brightness temperature and linear line bias. Lastly, accurate characterization of foregrounds and observational systematics are crucial in obtaining more realistic forecasts. 

\begin{acknowledgments}
It is our pleasure to thank David Alonso and Filippo Vernizzi for insightful discussions. A.M.D acknowledges funding from Tomalla Foundation for Research in Gravity. E.B.~ has received funding from the European Union’s Horizon 2020 research and innovation program under the Marie Skłodowska-Curie grant agreement No 754496.\\
\end{acknowledgments}

\appendix 
\vspace{-0.2in}
\section{Parameter constraints for a hypothetical staged LIM survey}\label{sec:app}
Tables \ref{tab:optsurv_1sig_cov} and \ref{tab:optsurv_1sig_effDE} show the 1-$\sigma$ parameter constraints for models described by a covariant approach and an effective parametrization, respectively. For each model, we show the constraints neglecting (top row) and including (bottom row) the interloper noise as optimistic and pessimistic cases. The number of spectrometer hours is representative of the timeline of the respective surveys, ranging from current generation surveys ($\sim 10^5$ spec-hours)  to those that can be envisioned in the next decade ($\sim 10^9$ spec-hours). The corresponding sky converge is obtained from optimization of survey strategy. Further details on survey specifications can be found in \cite{MoradinezhadDizgah:2021upg}. \\

\begin{table*}[htbp!]
\begin{center}
\hspace*{-3.5cm} \begin{tabular}{|c||c|c|c|c|c|c|c|c|c|}
\hline 
Spec & \multicolumn{2}{c|}{JBD} & \multicolumn{3}{c|}{Early-DE} & \multicolumn{4}{c|}{ShiftSymm-Horndeski}\tabularnewline
\cline{2-10} \cline{3-10} \cline{4-10} \cline{5-10} \cline{6-10} \cline{7-10} \cline{8-10} \cline{9-10} \cline{10-10} 
hours & $\omega_{\textrm{BD}}^{-1}$ & $\tilde{G}_{\textrm{eff}}$ & $f_{\textrm{EDE}}$ & $\log_{10}\left(z_{c}\right)$ & $\theta_{i}$ & $w_{0}$ & $w_{a}$ & $\hat{\alpha}_{\rm B}$ & $m$\tabularnewline
\hline 
\hline 
\multirow{2}{*}{$10^{5}$} & 0.00614  & 0.0953  & 0.0856  & 1.03  & 6.56  & 0.17  & 0.426  & 3.09  & 4.86 \tabularnewline
\cline{2-10} & 0.00848  & 0.142  & 0.117  & 1.45  & 9.16  & 0.303  & 0.967  & 5.48  & 8.91 \tabularnewline
\hline
\multirow{2}{*}{$10^{6}$} & 0.00277  & 0.0319  & 0.0404  & 0.345  & 2.01  & 0.0571  & 0.101  & 0.798  & 1.42 \tabularnewline
\cline{2-10} & 0.00388  & 0.0479  & 0.052  & 0.482  & 2.9  & 0.099  & 0.305  & 1.74  & 3.03 \tabularnewline
\hline
\multirow{2}{*}{$10^{7}$} & 0.000981  & 0.0106  & 0.0159  & 0.127  & 0.641  & 0.0223  & 0.0322  & 0.251  & 0.48 \tabularnewline
\cline{2-10} & 0.00151  & 0.0163  & 0.0242  & 0.18  & 0.927  & 0.0367  & 0.0972  & 0.556  & 0.99 \tabularnewline
\hline
\multirow{2}{*}{$10^{8}$} & 0.000338  & 0.00379  & 0.00555  & 0.0475  & 0.226  & 0.00847  & 0.0105  & 0.0814  & 0.184 \tabularnewline
\cline{2-10} & 0.000522  & 0.00569  & 0.00892  & 0.077  & 0.397  & 0.0139  & 0.0316  & 0.18  & 0.343 \tabularnewline
\hline
\multirow{2}{*}{$10^{9}$} & 0.000147  & 0.00178  & 0.00254  & 0.0229  & 0.123  & 0.00408  & 0.00354  & 0.0273  & 0.0792 \tabularnewline
\cline{2-10} & 0.000299  & 0.0034  & 0.0053  & 0.0535  & 0.308  & 0.00823  & 0.0168  & 0.0904  & 0.193 \tabularnewline
\hline
\end{tabular}
\end{center}
\caption{Potential stages of future mm-wave LIM experiments and the corresponding 1-$\sigma$ marginalized parameter constraints for models described by a covariant Lagrangian. All numbers include Planck $\Lambda$CDM priors. The fiducial values for the parameters are given in \S \ref{sec:th_MG_DE}. For each choice of spectrometer hour, the top row shows the constraints neglecting interlopers in the noise, while the bottom rows include the interlopers. See Table 2 of Ref.~\cite{MoradinezhadDizgah:2021upg} for approximate examples of the class of instrument required for each stage of the surveys.}
\label{tab:optsurv_1sig_cov}
\end{table*}
\vspace{-0.3in}
\begin{table*}[htbp!]
\begin{center}
\begin{tabular}{|c||c|c|c|c|c|c|c|c|c|c|c|c|}
\hline 
Spec & \multicolumn{2}{c|}{CPL} & \multicolumn{5}{c|}{Effective-DE, model1} & \multicolumn{5}{c|}{Effective-DE, model2}\tabularnewline
\cline{2-13} \cline{3-13} \cline{4-13} \cline{5-13} \cline{6-13} \cline{7-8} \cline{8-13} \cline{9-13} \cline{10-13} \cline{11-13} \cline{12-8} \cline{13-13} 
hours & $w_{0}$ & $w_{a}$ & $w_{0}$ & $w_{a}$ & $c_{M}^{(0)}$ & $c_{M}^{(1)}$ & $c_{M}^{(2)}$ & $w_{0}$ & $w_{a}$ & $c_{M}^{(0)}$ & $c_{M}^{(1)}$ & $c_{M}^{(2)}$\tabularnewline
\hline 
\hline 
\multirow{2}{*}{$10^{5}$} & 0.199  & 0.604  & 0.781  & 0.773  & 19.9  & 34.2  & 14.5  & 1.21  & 1.16  & 13.1  & 12.8  & 0.48 \tabularnewline
\cline{2-13} & 0.28  & 0.793  & 0.901  & 0.891  & 24.3  & 41.8  & 17.6  & 1.38  & 1.32  & 14.8  & 14.5  & 0.434 \tabularnewline
\hline
\multirow{2}{*}{$10^{6}$} & 0.0668  & 0.182  & 0.175  & 0.175  & 3.44  & 5.81  & 2.44  & 0.257  & 0.247  & 2.75  & 2.68  & 0.0625 \tabularnewline
\cline{2-13} & 0.0943  & 0.273  & 0.23  & 0.23  & 6.29  & 10.8  & 4.52  & 0.338  & 0.325  & 3.66  & 3.59  & 0.102 \tabularnewline
\hline
\multirow{2}{*}{$10^{7}$} & 0.027  & 0.0603  & 0.0551  & 0.0542  & 0.947  & 1.57  & 0.672  & 0.0791  & 0.0759  & 0.887  & 0.863  & 0.0163 \tabularnewline
\cline{2-13} & 0.0382  & 0.0908  & 0.0734  & 0.0723  & 1.96  & 3.38  & 1.43  & 0.106  & 0.101  & 1.18  & 1.15  & 0.032 \tabularnewline
\hline
\multirow{2}{*}{$10^{8}$} & 0.011  & 0.0239  & 0.0183  & 0.0178  & 0.296  & 0.499  & 0.229  & 0.0262  & 0.0251  & 0.302  & 0.293  & 0.0055 \tabularnewline
\cline{2-13} & 0.0157  & 0.0342  & 0.0248  & 0.0238  & 0.632  & 1.1  & 0.47  & 0.0356  & 0.0335  & 0.403  & 0.391  & 0.0105 \tabularnewline
\hline
\multirow{2}{*}{$10^{9}$} & 0.00537  & 0.0109  & 0.00592  & 0.00585  & 0.0941  & 0.16  & 0.0746  & 0.00853  & 0.00826  & 0.0985  & 0.0955  & 0.0018 \tabularnewline
\cline{2-13} & 0.00917  & 0.0189  & 0.00837  & 0.00813  & 0.215  & 0.4  & 0.159  & 0.0119  & 0.0112  & 0.141  & 0.137  & 0.00345 \tabularnewline
\hline
\end{tabular}
\end{center}
\caption{Same as Table \ref{tab:optsurv_1sig_cov}, but for models based on the effective description of DE.}
\label{tab:optsurv_1sig_effDE}
\end{table*}

\newpage
\bibliography{LIM_MG_DE}

\end{document}